\documentclass[PRB]{revtex4}
\usepackage{amssymb}
\usepackage{amsmath}
\usepackage{bbm}
\usepackage{graphicx}

\begin{document}

\title{Hybridization expansion impurity solver: General formulation and 
application to Kondo lattice and two-orbital models}
\author{Philipp Werner and Andrew J. Millis}
\affiliation{Columbia University, 538 West, 120th Street, New York, NY 10027, USA}

\date{August 30, 2006}

\hyphenation{}

\begin{abstract}
A recently developed continuous time solver based on an expansion in  hybridization about an exactly solved local limit is 
reformulated in a manner appropriate for general classes of quantum  impurity models
including spin exchange and pair hopping terms. The utility of the approach is demonstrated
via applications to the dynamical mean field theory of the Kondo lattice and two-orbital models.  
The algorithm can handle low temperatures and strong couplings without encountering a sign
problem.
\end{abstract}


\maketitle

\section{Introduction}

One of the fundamental challenges of theoretical condensed matter physics is the
accurate solution of quantum impurity
models. These, in general terms, consist of a 
Hamiltonian involving a finite number of states and a hybridization process which
allows particle exchange with one or more ``reservoirs" of particles. 
They are important both in their own right and as a crucial ingredient in the 
dynamical mean field (DMFT) \cite{Georges96} method of approximating the 
properties of interacting fermions on a lattice. Examples include the familiar 
Kondo and Anderson Hamiltonians and their generalization to multi-spin and multi-orbital cases,
as well as to the ``embedded plaquettes" used in the recently developed cluster 
extensions of dynamical mean field theory \cite{Jarrell, CDMFT, Fuhrmann04}.
 
Quantum impurity models  may be formulated as quantum field theories
in zero space and one time dimension, and the reduced dimensionality suggests that numerical approaches should be feasible.
However, up to now general quantum impurity models have to a large degree resisted numerical
attack. A special but conceptually crucial model, the one-orbital Anderson impurity model,
has been  studied in detail
but the techniques (the Hirsch-Fye discrete Hubbard-Stratonovich
transformation \cite{Hirsch86} and exact diagonalization \cite{Caffarel94}) 
which work relatively well in this case have proven difficult
to extend to  wider classes of models of physical interest. 

One issue is that the Hirsch-Fye method cannot easily be applied to 
models with interactions other than direct density-density couplings.
In particular, there is no good decoupling
for the  exchange and ``pair hopping" terms which are important
to the physics of partially filled d-levels. 
A scheme proposed by Sakai {\it et al.} \cite{Sakai04} has been used in some 
DMFT studies \cite{Arita05, Koga05}, but the method has a severe sign problem which
prevents calculations at low temperatures. 
Another issue with Hirsch-Fye and similar methods is time discretization, 
and in particular the fine grid spacing required to capture the 
short time behavior  of the Green function. The computational burden in Hirsch-Fye type methods 
grows as the cube of the (large) grid size, 
which must be increased
linearly with interaction strength and inverse temperature.
This severely restricts the accessible parameter range.   

The exact diagonalization method \cite{Caffarel94} represents the continuous density of states
of the reservoir by a small number of levels--but the number of levels required
scales linearly with the number of orbitals included while the computational burden
grows exponentially with the number of levels. This limits the applicability of the method to models 
with a small number of orbitals, although some results have been presented for three orbital models \cite{Capone02} and four-site clusters \cite{Kyung06}.  

Recently, a new class of impurity solvers has been 
developed \cite{Rubtsov05, Werner06}, based  on the
stochastic evaluation of a diagrammatic expansion of the partition function.
Two complimentary approaches are possible, based on 
a weak-coupling
expansion in powers of the coupling constants \cite{Rubtsov05} or  
an 
expansion in powers of the impurity-bath mixing \cite{Werner06}. These algorithms, which 
require neither auxiliary fields nor a time discretization, 
have been shown to provide considerable improvements over the Hirsch-Fye
method for the one-orbital Anderson model.  The weak coupling approach has
also been successfully applied to more complicated models \cite{Savkin05}, and an interesting hybrid
scheme involving a Hirsch-Fye decoupling of density channel interactions and an expansion
in exchange interactions has very recently been applied to multiorbital models
\cite{Rombouts99, Sakai06}.

In Ref.~\cite{Werner06} we have demonstrated the usefulness of the hybridization 
expansion approach for the single site Hubbard model. Its power relies on the fact that
the order of perturbation which is needed decreases as the interaction strength
increases. The algorithm was found to 
allow access to extremely low temperatures, even in the presence of strong interactions.
But the formulation given in Ref.~\cite{Werner06} was specific to models (such as the Hubbard model)
with only density-density interactions. In this paper we present a matrix formulation
which generalizes  the 
method to wide classes of impurity models.
To demonstrate 
the power of the hybridization expansion approach we use it to calculate physical properties
of the dynamical mean field approximation to the Kondo lattice model 
(for which only very few DMFT calculations have been attempted)
and the multiorbital Anderson model.

\section{Formalism}


A general impurity model contains fermions labeled by quantum numbers 
$a=1,\ldots,N$ (denoting for example site, spin and orbital indices), interacting
with each other,  coupled to local degrees of freedom $\textbf{T}$ (representing for example
spin or phonon fields) and hybridized with ``bath" fermions. The latter have a 
continuous density of states which we parametrize by ``momentum" $p$.
It is convenient to assemble the fermion fields and the bath fermions into $N$-component 
spinors $\psi$ and $b$, respectively. The general Hamiltonian is then
\begin{eqnarray}
H &=& H_\text{loc}+H_\text{bath}+H_\text{hyb}+H_\text{hyb}^\dagger,\label{H}
\end{eqnarray}
with
\begin{eqnarray}
H_\text{loc} &=& \psi^\dagger \textbf{Q} \psi \cdot \textbf{T}+h.c.+H_T+\sum_{a,b,c,d}U^{abcd}\psi_b^\dagger\psi_c^\dagger\psi_c\psi_d,\label{H_loc}\\
H_\text{bath} &=& \sum_p \epsilon_p b^\dagger_p b_p,\label{H_bath}\\
H_\text{hyb} &=& \sum_p \psi \textbf{V}_p b^\dagger_p.\label{H_hyb}
\end{eqnarray}
We have assumed here that the fermion-fermion interaction is of the conventional four-fermion
type, but the extension to more general forms is immediate. Similarly,
we have assumed a bilinear coupling (specified by some matrix $\textbf{Q}$)  
between the local fermions and the spin and lattice
degrees of freedom represented by ${\bf T}$, but more general interactions are easily
included.     

The ``bath" fermions are assumed to be orthogonal and to have free fermion
correlations while
$\textbf{V}$ is an $N\times N$ hybridization matrix, which has to be determined in a 
self-consistent manner. 
In the impurity models known to us it is possible to find a representation in which
$H_\text{bath}$ and  $\textbf{V}$ are simultaneously diagonal, that is
\begin{equation}
H_\text{hyb}+H_\text{bath}=\sum_{a,p} \psi_a V^a_p {b^a_p}^\dagger+\sum_{a,p}\epsilon_p{b^a_p}^\dagger b^a_p 
=\sum_a H_\text{hyb}^a+\sum_a H_\text{bath}^a,\label{H_hyb_diag}
\end{equation}
and we make this assumption throughout the rest of this paper.

The impurity model partition function $Z$ may then be expressed  as
\begin{eqnarray}
Z=Z_\text{bath}Tr_\psi \left\langle T_\tau e^{-\int_0^\beta d\tau H_\text{loc}(\tau)+H_\text{bath}(\tau)+\sum_a (H^a_\text{hyb}(\tau)+H^{a \dagger}_\text{hyb}(\tau)) }\right\rangle_b,\label{Z}
\end{eqnarray}
with $Z_\text{bath}=Tr_b e^{-\beta H_\text{bath}}$ and $\langle.\rangle_b=Tr_b[.]/Z_\text{bath}$. 

We expand Eq.~(\ref{Z}) in the hybridizations $\psi_aV^a_pb^{a\dagger}_p$ and 
$b^a_p{V^{a}_p}^\star\psi_a^\dagger$. Each term in the expansion must have 
the same number of $\psi_a$ and $\psi_a^\dagger$ operators, so
\begin{eqnarray}
Z &=& Z_\text{bath}Tr_\psi \Big\langle T_\tau e^{-\int_0^\beta d\tau H_\text{loc}(\tau)+H_\text{bath}(\tau)}\prod_a \sum_{k_a} Z_{k_a}\Big\rangle_b,\\
Z_{k_a} &=& \sum_{p_1,...,p_{k_a}}\sum_{p'_1,...,p'_{k_a}}V_{p_1}^a {V_{p'_1}^a}^{\!\star} ...V_{p_{k_a}}^a {V_{p'_{k_a}}^a}^{\!\!\!\!\star} \int_0^\beta d\tau_1 \int_{\tau_1}^{\beta}d\tau_2 \dots \int_{\tau_{k_a-1}}^\beta d\tau_{k_a}\int_0^\beta d\tau'_1 \int_{\tau'_1}^{\beta}d\tau'_2 \ldots \int_{\tau'_{k_a-1}}^\beta d\tau'_{k_a} \nonumber\\
&&\times \psi_a(\tau_1)b^{a\dagger}_{p_1}(\tau_1)b^a_{p'_1}(\tau'_1)\psi^\dagger_a(\tau'_1)
\psi_a(\tau_2)b^{a\dagger}_{p_2}(\tau_2)b^a_{p'_2}(\tau'_2)\psi^\dagger_a(\tau'_2)\ldots
\psi_a(\tau_{k_a})b^{a\dagger}_{p_{k_a}}(\tau_{k_a})b^a_{p'_{k_a}}(\tau'_{k_a})\psi^\dagger_a(\tau'_{k_a}),\label{Z_k_a}
\end{eqnarray}
where we have used the $1/k_a!$ in time ordering the $\psi$'s and $\psi^\dagger$'s. 
We now take the expectation value over the bath states. The unprimed and primed 
$p$-indices must always occur in pairs $p_i=p'_j\equiv p$ and tracing over the 
bath states  thus yields a factor 
$|V^a_p|^2 e^{-\epsilon_p(\tau'_j-\tau_i)}/(e^{-\beta \epsilon_p}+1)$ if $\tau'_j>\tau_i$ 
and  $|V^a_p|^2 e^{-\epsilon_p(\tau'_j-\tau_i+\beta)}/(e^{-\beta \epsilon_p}+1)$ 
if $\tau'_j<\tau_i$. By defining the \textit{hybridization function} $F_a(\tau)$ as
\begin{eqnarray}
F_a(\tau)=\left\{ \begin{array}{ll} \sum_p  |V^a_p|^2 e^{-\epsilon_p(\beta-\tau)}/(e^{-\beta\epsilon_p}+1)& \tau>0 \\
                                                          \sum_p -|V^a_p|^2 e^{-\epsilon_p(-\tau)}/(e^{-\beta\epsilon_p}+1)& \tau<0 
                                     \end{array} \right.,
                                     \label{F_a}
\end{eqnarray}
the expectation value of the $b$-operators can be expressed as the determinant of a matrix $M_a^{-1}$ with elements
\begin{equation}
M_a^{-1}(i,j)=F_a(\tau_i-\tau'_j).\label{M_a_inv}
\end{equation} 

Note that
\begin{equation}
F(-i\omega_n)=\int d\tau e^{-i\omega_n\tau}F(\tau)=\int d\omega \sum_p |V_p|^2 \frac{\delta(\omega-\epsilon_p)}{i\omega_n-\omega},
\end{equation}
so that the hybridization 
functions $F$ are the same as those defined in Ref.~\cite{Werner06} 
and are related
to the conventionally defined ``Weiss function"  ${\cal G}_0^{-1}$ \cite{Georges96}
by $F(-i\omega_n)=i\omega_n+\mu-{\cal G}_0^{-1}(i\omega_n)$. 

The partition function finally becomes
\begin{eqnarray}
Z &=& Z_\text{bath}Tr_\psi \left[T_\tau e^{-\int_0^\beta d\tau H_\text{loc}(\tau)}\prod_a \sum_{k_a} \tilde Z_{k_a}\right]s_{T_\tau},\label{Z_final}\\
\tilde Z_{k_a} &=& \int_0^\beta d\tau_1 \int_{\tau_1}^{\beta}d\tau_2 \dots \int_{\tau_{k_a-1}}^\beta d\tau_{k_a}\int_0^\beta d\tau'_1 \int_{\tau'_1}^{\beta}d\tau'_2 \ldots \int_{\tau'_{k_a-1}}^\beta d\tau'_{k_a} \det(M_a^{-1})s_a\nonumber\\
&&\times \psi_a(\tau_1)\psi^\dagger_a(\tau'_1)
\psi_a(\tau_2)\psi^\dagger_a(\tau'_2)\ldots
\psi_a(\tau_{k_a})\psi^\dagger_a(\tau'_{k_a}),\label{Z_psi}
\end{eqnarray}
with $s_a$ a sign 
determined by the signature of the permutation which 
permutes the $a$-flavored field operators from their time-ordered sequence (smallest $\tau$ shifted to the right) into the alternating order 
$\psi_a(\tau_1)\psi^\dagger_a(\tau'_1)\psi_a(\tau_2)\psi^\dagger_a(\tau'_2)\ldots$, and $s_{T_\tau}$ 
compensating for an eventual sign change produced by the time ordering of all the operators.
The sign factor
$s_a$ arises from the $\beta$-antiperiodic definition of $F_a$ (Eq.~(\ref{F_a})) and is the generalization of the signs denoted $\delta_{\tau_1^s}^{\tau_k^e}$ in Ref.~\cite{Werner06}.
The sign $s_{T_\tau}$ is merely a consequence of the notation in Eq.~(\ref{Z_final}), where we grouped together all of the operators
corresponding to a given flavor $a$. If all the $\psi$ operators, irrespective of flavor, are placed in the order in which they
occur, there is no additional sign.

\section{Monte Carlo procedure}

\begin{figure}[t]
\centering
\includegraphics [angle=0, width= 7.5cm] {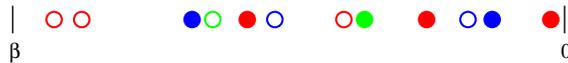}
\caption{(Color online) Every Monte Carlo configuration can be represented by a sequence of 
operators on the time interval $0\le\tau<\beta$ (we let time run from right to left to be consistent with the time ordering convention). Different colors correspond to different flavors, 
while full (empty) circles represent creation (annihilation) operators. The Monte Carlo 
moves consist of random insertions or deletions of pairs of operators in the different channels.}
\label{general}
\end{figure}

Eqs.~(\ref{Z_final}) and (\ref{Z_psi}) show that the partition function may be expressed as a 
sum over configurations consisting of $2n=2\sum_a k_a$ operators $\{O_i(\tau_i)\}_{0\le \tau_1<\tau_2<\ldots<\tau_{2n}<\beta}$. Of these operators, 
$k_a$ are creation operators $\psi^\dagger_a$ and another $k_a$ are destruction operators 
$\psi_a$ and they are connected in all possible ways by hybridization functions 
$F_a$ (this is the interpretation of the determinant). Sandwiched in between 
the $O$'s are time evolution operators $K_\text{loc}$, defined as
\begin{equation}
K_\text{loc}(\tau) = e^{-H_\text{loc}\tau}.\label{K_loc}
\end{equation} 
A typical configuration can thus be illustrated by a sequence of dots on an 
interval $[0,\beta)$ representing imaginary time (see Fig.~\ref{general}). Each color 
corresponds to a different flavor $a$, while full and empty dots represent 
creation and annihilation operators, respectively.
The weight of such a configuration is given by 
\begin{equation}
w(\{O_i(\tau_i)\})= Tr \left[ K_\text{loc}(\beta-\tau_{2n}) O_{2n} (\tau_{2n}) \ldots O_2(\tau_2) K_\text{loc}(\tau_2-\tau_1) O_1(\tau_1) K_\text{loc}(\tau_1) \right] d\tau_1\ldots d\tau_{2n} \prod_a (\det M_a^{-1})s_a.\label{weight}
\end{equation} 

A Monte Carlo procedure which samples the whole configuration space is 
obtained by randomly inserting and removing pairs of operators in the $a$ 
channel ($a=1,\ldots,N$), or changing their position on the time interval. 
The detailed balance condition for insertion/removal of a pair in channel $a$ reads
\begin{equation}
\frac{p(\{O\}_{2n})\rightarrow p(\{\tilde O\}_{2n+2})}{p(\{\tilde O\}_{2n+2})\rightarrow p(\{O\}_{2n})}=\frac{\beta^2}{(k_a+1)^2}
\frac{Tr [ K_\text{loc}(\beta-\tilde \tau_{2n+2}) \tilde O_{2n+2} (\tilde \tau_{2n+2}) \ldots \tilde O_1(\tilde \tau_1) K_\text{loc}(\tilde \tau_1)]}
{Tr [ K_\text{loc}(\beta-\tau_{2n}) O_{2n} (\tau_{2n}) \ldots O_1( \tau_1) K_\text{loc}(\tau_1)]}\frac{\det \tilde M_a^{-1}\tilde s_a}{\det {M}_a^{-1}s_a}, \label{detailed_balance}
\end{equation}
and can be satisfied for example by using the Metropolis algorithm. 
In each update, it is therefore necessary to compute
both
the determinant of the new 
$F_a$-matrix, $\det M_a^{-1}$, and the trace of the new sequence of field operators and 
propagators. This latter task is simplified by writing all the operators in the eigenbasis of $H_\text{loc}$. 

In the simulation, one actually stores and manipulates $M_a$, the inverse of the matrix 
defined in Eq.~(\ref{M_a_inv}). Fast matrix updates, similar to the ones detailed in 
Ref.~\cite{Rubtsov05} allow to compute the new $M_a$ in a time $O(k_a^2)$. The 
elements of this matrix also yield the measurement values for the Green function 
$G_a$ at the time intervals given by the operator positions ($\tau_i$ for annihilation and 
$\tau'_j$ for creation operators), 
\begin{align}
G_a(\tau) &= \Big\langle\frac{1}{\beta}\sum_{i,j=1}^{k_a} M_a(j,i)\Delta(\tau, \tau_i-\tau'_j)\Big\rangle,\label{Green}\\
\Delta(\tau, \tau') &= \left\{ \begin{array}{ll} \delta(\tau-\tau') & \tau'>0 \\
                                                             -\delta(\tau-\tau'-\beta) & \tau'<0 
                                     \end{array} \right..\label{Delta}\hspace{5mm}
\end{align}
Angular brackets denote the Monte Carlo average. Other observables can be 
measured by computing a trace. For example, the mean particle number can be obtained as
\begin{equation}
n=\Big \langle \frac{Tr [ K_\text{loc}(\beta-\tau_{2n}) O_{2n} (\tau_{2n}) \ldots O_1( \tau_1) K_\text{loc}(\tau_1) \hat n ]}{Tr [ K_\text{loc}(\beta-\tau_{2n}) O_{2n} (\tau_{2n}) \ldots O_1( \tau_1) K_\text{loc}(\tau_1)]} \Big\rangle,
\end{equation}
where $\hat n$ is the number operator.

A computationally expensive part of this procedure is the evaluation of the trace in 
the acceptance rate of Monte Carlo moves. In general, there are certain combinations 
of operators which always yield a zero trace and checking these conditions beforehand 
allows one to avoid unnecessary computations of the trace. 

\begin{figure}[t]
\centering
\includegraphics [angle=0, width=8.5cm] {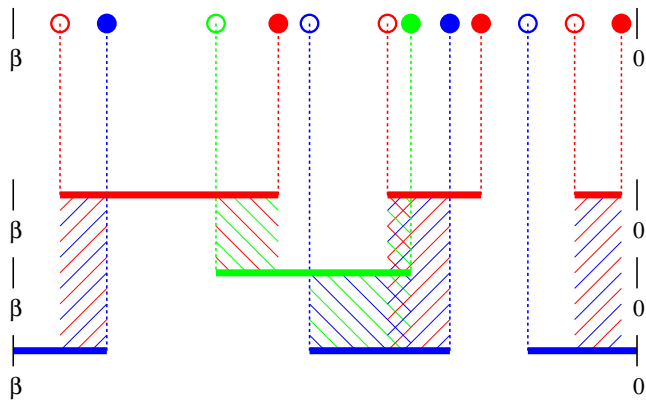}
\caption{(Color online) If the creation and annihilation operators for each flavor must occur in alternating order, as is the case for 
models without exchange and pair-hopping, then it is convenient to represent the 
configurations with non-zero trace by collections of segments. The weight of a 
segment configuration is determined by the length of the segments and the 
overlap between segments of different flavors (indicated by the hashed regions).}
\label{segment}
\end{figure}

Models which do not contain exchange or ``pair-hopping" processes,
so that $H_\text{loc}$ and the $\psi$-operators are  diagonal in the flavor indices $a$, constitute
a special case. For these models,  
the creation and annihilation operators for each flavor must occur in alternating order and as
shown in Ref.~\cite{Werner06}  the ``segment" representation, illustrated in
Fig.~\ref{segment}, is an efficient way of specifiying all the 
configurations of non-zero trace.  In this scheme, 
configurations are represented as collections 
of segments (one collection for each flavor), whose start and end points coincide with 
the positions of the creation and annihilation operators.  The weight of a configuration can 
be expressed in terms of the lengths of the segments and the overlaps between 
segments of different flavors. 

Care must be taken to prevent the system from being trapped in a state which breaks a symmetry of $H_\text{loc}$ when it should not be. For studying paramagnetic (para-orbital) phases, averaging the Green functions is sufficient. To study
broken symmetry phases, 
the Green functions corresponding to different spin (orbital) states must be allowed to evolve independently and to obtain a symmetry unbroken state (e. g. above some critical temperature) it is then important that the Monte Carlo sampling explores the whole configuration space. To avoid un-physical trapping, we introduce ``swap"-moves, which exchange the operators  corresponding for example to up- and down-spins in a given orbital. Because the calculation of the new $M_a$-matrices requires explicit matrix inversions, 
which are $O(k_a^3)$, swap-moves are costly, but a relatively small number of attempts is 
enough to assure an ergodic sampling.

\subsection{Self consistency and ordered phases}

The preceding sections described the solution of an impurity model specified by a local Hamiltonian and hybridization functions. In dynamical mean field theory, the hybridization functions are fixed by a self-consistency condition relating the impurity model Green function (\ref{Green}) to the implied lattice Green function. The precise form of the equation depends on the specific dynamical mean field equation chosen, so a general equation cannot be written here. A crucial point is that the
information concerning symmetry breaking is carried by the hybridization functions $F$ and 
enters the problem via the self-consistency condition. $H_\text{loc}$ (and therefore the matrix forms
of the creation and time evolution operators) retain their symmetry unbroken form.  

In this paper we use semicircular densities of states with (possibly orbital dependent)  
full bandwidths $4t_a$. The self-consistency condition for 
translationally-invariant states,
including both paramagnetic states and states with ferromagnetic or ferro-orbital order  is (the $-\tau$ follows from the definition of $F_a$ in Eq.~(\ref{F_a}))
\begin{equation}
F_a(\tau) = t_a^2 G_a(-\tau). \label{self_consistency}
\end{equation}

States with a broken translational invariance may also be studied. For example, for
bipartite lattices with simple two-sublattice Neel order or (in the case of the models
with two-fold orbital degeneracy) two-sublattice orbital order, the condition becomes
\begin{equation}
F_a(\tau) = t_a^2 G_{\bar a}(-\tau), \label{self_consistency_anti}
\end{equation}
where $\bar a$ denotes the opposite spin or the complementary orbital

In subsequent sections we illustrate the formalism via study of two models in which exchange interactions play an important role: the Kondo lattice model and the  Hubbard model 
with a two-fold orbital degeneracy.

\section{Application I: Kondo Lattice}
\subsection{Overview}

In the Kondo lattice model, a local spin-1/2 degree of freedom, $\textbf{S}$, couples via a coupling constant
$J$ which may be either negative (``ferromagnetic") or positive (``antiferromagnetic") 
to electrons which reside in a single orbital, so that Eq.~(\ref{H_loc}) becomes
\begin{equation}
H_\text{loc}=-\mu\sum_a \psi_a^\dagger \psi_a+J\vec{\bf S}\cdot \frac{1}{2}\psi_a^\dagger \vec \sigma_{a b}\psi_b.\label{H_kondo}
\end{equation}

The Kondo  impurity model, i.e. a single spin subject to a Hamiltonian $H_\text{Kondo}+H_\text{bath}$ with
$H_\text{bath}$ fixed  (no self consistency equation) and characterized by
a constant density of states $\rho$ near the fermi level, has been extensively studied.
The physics  exhibits a profound dependence on the sign
of the exchange constant $J$: for ferromagnetic $J$ the coupling scales 
asymptotically to zero
according to 
\begin{equation}
J_\text{eff}(\beta) \sim \frac{\rho J}{1+\rho J \ln (\beta/\rho)},
\label{RG}
\end{equation}
($J_\text{eff}(\omega) \sim 1/\ln(\omega)$) so that the asymptotic low temperature and low frequency
behavior is that of free moments decoupled from the conduction electrons.
On the other hand, for antiferromagnetic
sign the problem scales to strong coupling, leading to the formation of a Kondo resonance
and the dissolution of the spin
into the bath of conduction electrons. 

Less is known about the lattice problem.  We summarize here some results which
are relevant to the half-filled case studied in this paper.  For a {\em classical} spin the 
sign of $J$ is irrelevant and for a bipartite lattice and particle-hole-symmetric dispersion 
the ground state is an antiferromagnetic insulator for all $J$ \cite{Chattopadhyay01}. 
The paramagnetic phase  of the classical model is characterized by disordered spins, 
and may be an insulator at large $J$ or a metal at small $J$. In the metallic phase the
spin disorder implies a nonvanishing scattering rate at the fermi level. 

For $S=1/2$ quantum spins, fewer results have been presented. It is generally believed
that the half-filled, bipartite antiferromagnetic Kondo lattice exhibits a large-$J$ Kondo insulator phase
(the lattice version of the Kondo singlet behavior) whereas for smaller $J$ a phase transition
to an antiferromagnet occurs \cite{Doniach77,kondo_gap}. For the ferromagnetic
side even less is known.
A very recent study of the ferromagnetic Kondo lattice
model at $n\neq 1$, based on the ``equation of motion approach" 
which does not capture the Kondo scaling, reports a transition from a 
ferromagnetic to a paramagnetic state with increasing doping \cite{Kienert06}.

\subsection{Formalism}

We now turn to the specifics of the  solution of this problem using the new method. 
$H_\text{loc}$ is diagonal in the basis of total particle number, total spin and 
$z$-component of total spin. If the 
particle number is 0 or 2, then the spin state is just the state of the local moment, if 
the number is 1, the spin state is singlet ($S$) or triplet ($T_{m_z}$) with given $m_z$. 
The eigenstates may thus be labeled as shown in Tab.~\ref{eigenstates_kondo},
\begin{table}[h]
\begin{tabular}{ll}
Eigenstates \hspace{10mm}\mbox& Energy\\
&\\
$|1\rangle = |0,\uparrow\rangle$ & 0 \vspace{1mm}\\
$|2\rangle = |0,\downarrow\rangle$ & 0 \vspace{1mm}\\
$|3\rangle = |1,S\rangle$ & $-\frac{3}{4}J-\mu$ \vspace{1mm}\\
$|4\rangle = |1,T_1\rangle$ & $\frac{1}{4}J-\mu$ \vspace{1mm}\\
$|5\rangle = |1,T_0\rangle$ & $\frac{1}{4}J-\mu$ \vspace{1mm}\\
$|6\rangle = |1,T_{-1}\rangle$ & $\frac{1}{4}J-\mu$ \vspace{1mm}\\
$|7\rangle = |2,\uparrow\rangle$ & $-2\mu$ \vspace{1mm}\\
$|8\rangle = |2,\downarrow\rangle$ & $-2\mu$ \vspace{1mm}\\
\end{tabular}
\caption{Eigenstates and eigenenergies for the local part of the Kondo lattice hamiltonian. The first entry labels the number of electrons and the second entry the spin state: either impurity spin $\uparrow, \downarrow$ if the number  of electrons is $0$ or $2$
or the total spin $S$ (singlet) $T_m$ (triplet with $m_z=m$) if $n=1$.}
\label{eigenstates_kondo}
\end{table}
where the first entry is the number of electrons and the second entry refers to the spin state. The singlet state is defined as $S=\frac{1}{\sqrt{2}}(|\!\uparrow,\downarrow\rangle - |\!\downarrow,\uparrow\rangle)$, with the first entry the conduction electron and the second entry the local moment spin direction. In this basis, the time evolution operator is diagonal, $K(\tau)|n\rangle=\exp(-E_n\tau)|n\rangle$, with eigenenergies $E_n$ listed in Tab.~\ref{eigenstates_kondo}. The creation operators for spin up and down become the sparse matrices
\begin{equation}
\psi_\uparrow^\dagger =\left( \begin{array}{cccccccc} 	0\hspace{0mm} & 0\hspace{0mm}  & 0\hspace{0mm}  & 0\hspace{0mm}  & 0\hspace{0mm}  & 0\hspace{0mm}  & 0\hspace{0mm}  & 0\hspace{0mm}  \\
										0 & 0 & 0 & 0 & 0 & 0 & 0 & 0 \\
										0 & \frac{1}{\sqrt{2}} & 0 & 0 & 0 & 0 & 0 & 0 \\
										1 & 0 & 0 & 0 & 0 & 0 & 0 & 0 \\
										0 & \frac{1}{\sqrt{2}} & 0 & 0 & 0 & 0 & 0 & 0 \\
										0 & 0 & 0 & 0 & 0 & 0 & 0 & 0 \\
										0 & 0 & \frac{-1}{\sqrt{2}} & 0 & \frac{1}{\sqrt{2}} & 0 & 0 & 0 \\
										0 & 0 & 0 & 0 & 0 & 1 & 0 & 0 \\
                                     \end{array} \right),
                                     \hspace{10mm}
\psi_\downarrow^\dagger =\left( \begin{array}{cccccccc} 	0\hspace{0mm}  & 0\hspace{0mm}  & 0\hspace{0mm}  & 0\hspace{0mm}  & 0\hspace{0mm}  & 0\hspace{0mm}  & 0\hspace{0mm}  & 0\hspace{0mm}  \\
										0 & 0 & 0 & 0 & 0 & 0 & 0 & 0 \\
										\frac{-1}{\sqrt{2}} & 0 & 0 & 0 & 0 & 0 & 0 & 0 \\
										0 & 0 & 0 & 0 & 0 & 0 & 0 & 0 \\
										\frac{1}{\sqrt{2}} & 0 & 0 & 0 & 0 & 0 & 0 & 0 \\
										0 & 1 & 0 & 0 & 0 & 0 & 0 & 0 \\
										0 & 0 & 0 & 1 & 0 & 0 & 0 & 0 \\
										0 & 0 & \frac{1}{\sqrt{2}} & 0 & \frac{1}{\sqrt{2}} & 0 & 0 & 0 \\
                                     \end{array} \right).    
\label{psi_kondo}                                                                   
\end{equation}
With these operators, the sampling then proceeds as described in the previous sections.

An important issue for simulations of interacting fermion problems is the sign of the different contributions
to the partition sum. For the Hubbard model  we noted the empirical absence 
of a sign problem in Ref.~\cite{Werner06}. This absence of sign is not unexpected: the density-density interaction is essentially classical (no exchange) and other simulation methods do not give rise to a sign 
problem in this case. One might expect the situation in the Kondo lattice model
to be worse, because it contains explicit exchange processes. Indeed
as can be seen from Eq.~(\ref{psi_kondo}), the matrix elements for transitions into or out of 
singlet states can be negative. However, since these negative matrix elements always occur in pairs, 
the trace in Eq.~(\ref{weight}) is not a source of sign problems. Negative determinants of the 
$F$-matrices could in principle lead to negative weights (note that the exchange processes
lead to operator orderings not found in the Hubbard model). 
Surprisingly, we do not find a sign problem in our simulations of the Kondo lattice either. For the
 parameters used in most of this investigation, $\beta t=50$ or $100$, $-10\le J/t\le 1$ 
 and densities per spin $n\le 0.98$, the average sign is~1. Configurations with negative weight exist,
 but  contribute negligibly little to the partition sum and are hence not generated. On some occasions, we measured average signs which differed from one in the sixth or seventh decimal place, but the converged solutions were usually not affected in any way by negative-weight contributions.
 
\begin{figure}[t]
\centering
\includegraphics [angle=-90, width=8.5cm] {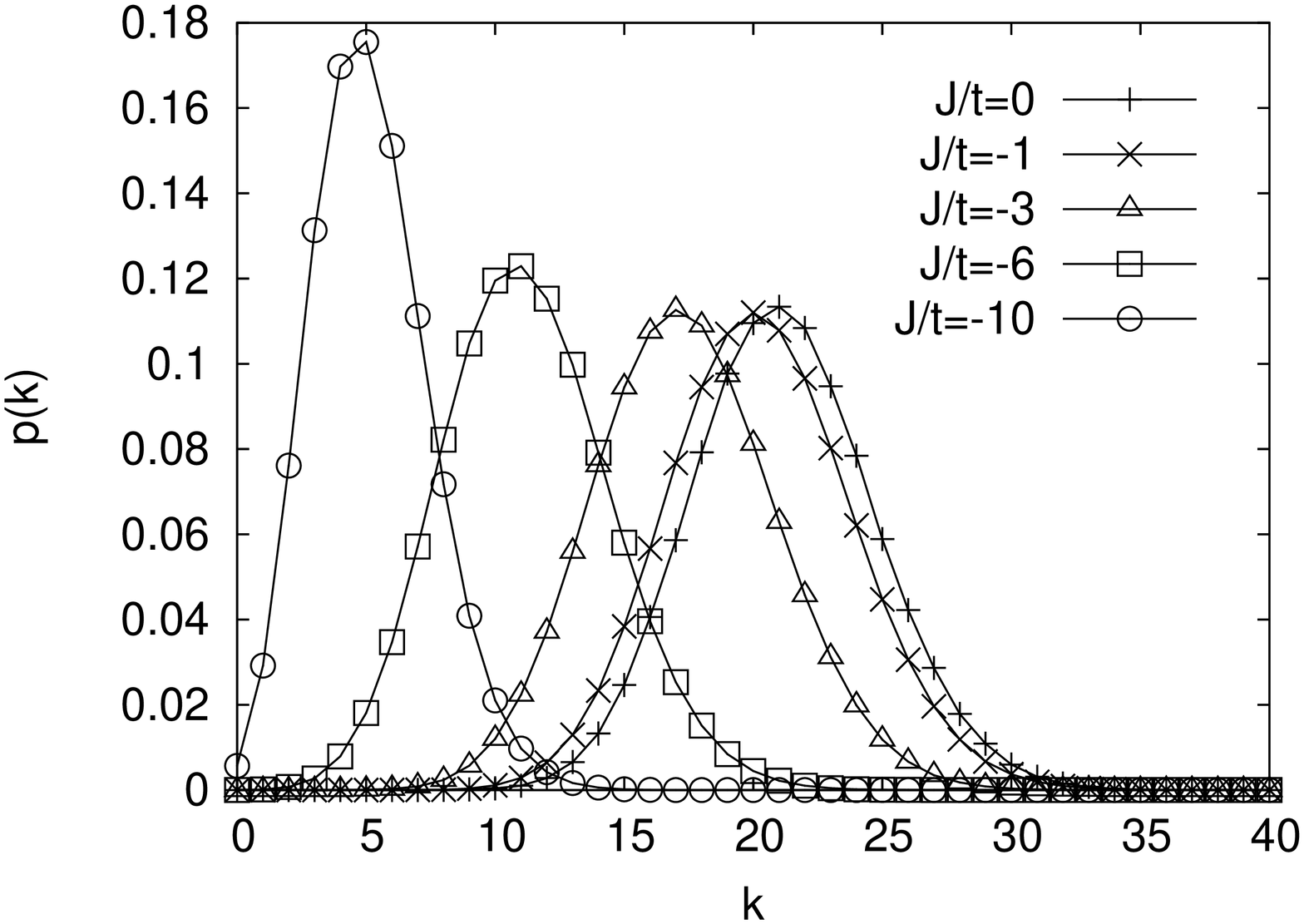}
\includegraphics [angle=-90, width=8.5cm] {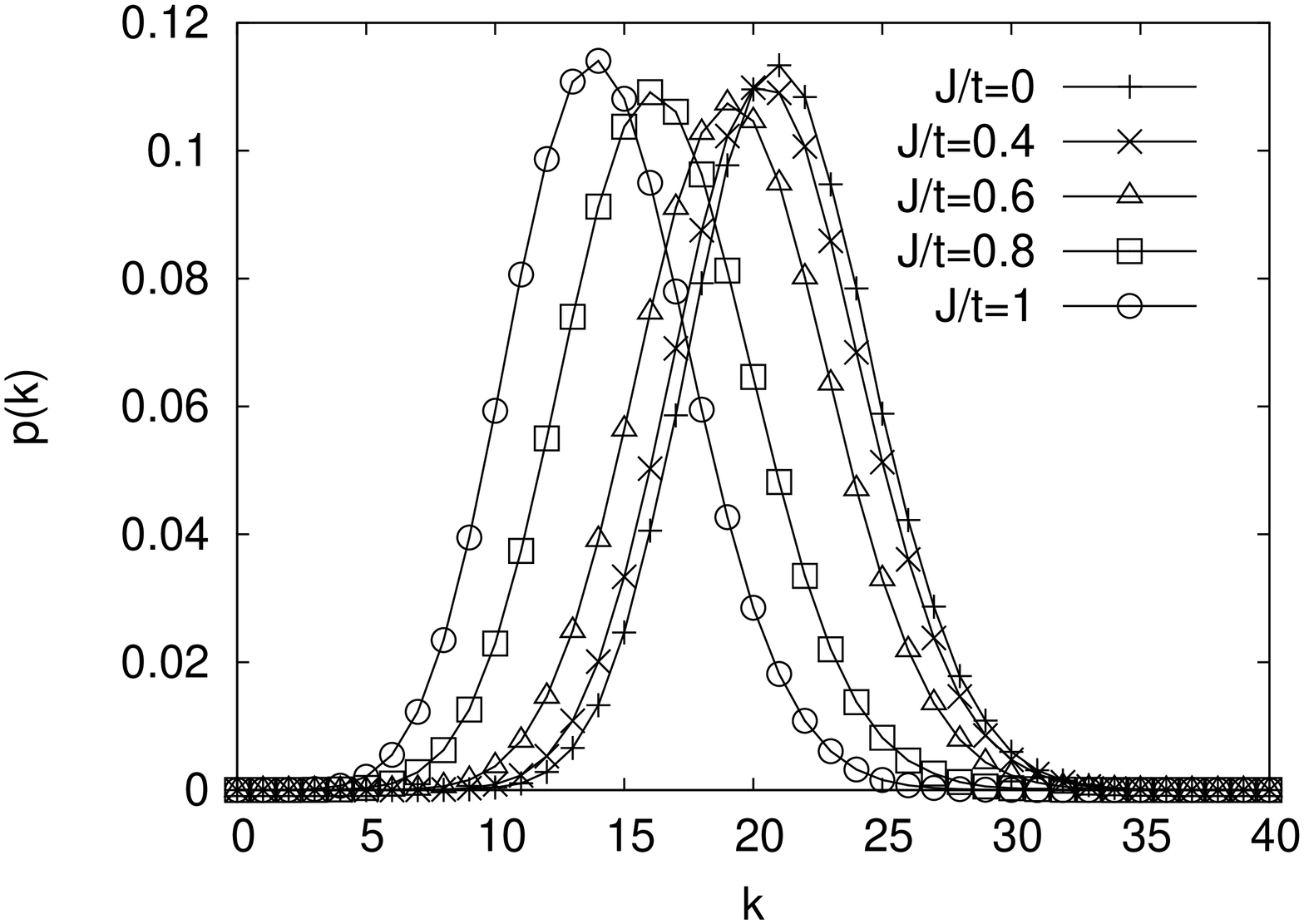}
\caption{Distribution of perturbation orders $p(k)\equiv p(k_\uparrow)=p(k_\downarrow)$ for the  ferromagnetic (left panel) and anti-ferromagnetic (right panel) kondo lattice models 
at half filling and inverse temperature $\beta t=50$. 
Note the different $J$-ranges in the two panels.
The mean perturbation order shifts lower  as the coupling magnitude $|J|$ is increased. 
For antiferromagnetic coupling, this effect is much more pronounced. }
\label{order_kondo}
\end{figure}

A particularly attractive feature of the hybridization expansion approach is the fact that 
stronger interactions lead to a lower perturbation order, independent of the sign of $J$. In Fig.~\ref{order_kondo} we plot the 
distribution of perturbation orders $p(k_\uparrow)=p(k_\downarrow)$ corresponding to the converged solutions for different values of $J/t$ and $\beta t=50$. While the 
distribution shifts in a way which is comparable to the one observed in the Hubbard model \cite{Werner06} for $J<0$, the effect is even more pronounced for $J>0$. For all parameter values considered in this study, the perturbation order remains reasonably low and thus allows an efficient Monte Carlo sampling.

\subsection{Paramagnetic phase}

\subsubsection{Overview and classical limit}

In this subsection we consider the behavior of the model in the paramagnetic phase
(with magnetism suppressed 
by symmetrization of the Green function). For orientation, we first briefly discuss the physics of the 
classical core-spin model, in the paramagnetic phase.  
As noted above, in the classical model
the sign of the exchange is irrelevant, and in the paramagnetic phase
the spins are disordered and provide a static spin-dependent scattering potential
for the electrons. 
In the dynamical mean field approximation to the classical spin model one finds that the
self energy is $\Sigma(\omega)=J_\text{eff}^2/{\cal G}_0^{-1}$
with $2J_\text{eff}=J/2$ the up-down energy splitting arising from the diagonal part of the 
exchange term in Eq.~(\ref{H_kondo}) and the mean 
field function ${\cal G}_0^{-1}$ is given by
\begin{equation}
{\cal G}_0^{-1}(\omega)=\omega+\mu-t^2\frac{{\cal G}_0^{-1}}{\left({\cal G}_0^{-1}\right)^2-J_\text{eff}^2}.
\label{classicalKondo}
\end{equation}
At half filling ($\mu=0$) and at the fermi level $(\omega=0)$ 
this equation has two solutions:
\begin{eqnarray}
{\cal G}_0^{-1}&=&0,
\label{Ginsulating}
\\
{\cal G}_0^{-1}&=&i\sqrt{t^2-J_\text{eff}^2}.
\label{G0classicalkondo}
\end{eqnarray}
Eq.~(\ref{G0classicalkondo}) describes a metal ($\Im m G \neq 0$ at the fermi level) with
a self energy
\begin{eqnarray}
\Sigma&=&-i\frac{J_\text{eff}^2}{\sqrt{t^2-J_\text{eff}^2}}
\label{Sigmaclassicalkondo}
\end{eqnarray}
which has a non-vanishing imaginary part, corresponding to scattering of electrons off the 
static spins. As $J_\text{eff} \rightarrow t$ the fermi level density of states vanishes and the scattering
rate diverges. For $\left|J_\text{eff}\right|>t$ the relevant solution is that of
Eq.~(\ref{Ginsulating}), which is the $\omega\rightarrow 0$ limit of the expected insulating result ${\cal G}_0^{-1}\sim \omega$ (describing an insulator with $\Sigma \sim \frac{J^2}{i\omega}$).

In the rest of this section we present results for the quantum model, where the physics
depends on the sign of $J$.

\subsubsection{Ferromagnetic $J$}

\begin{figure}[t]
\centering
\includegraphics [angle=-90, width=8.5cm] {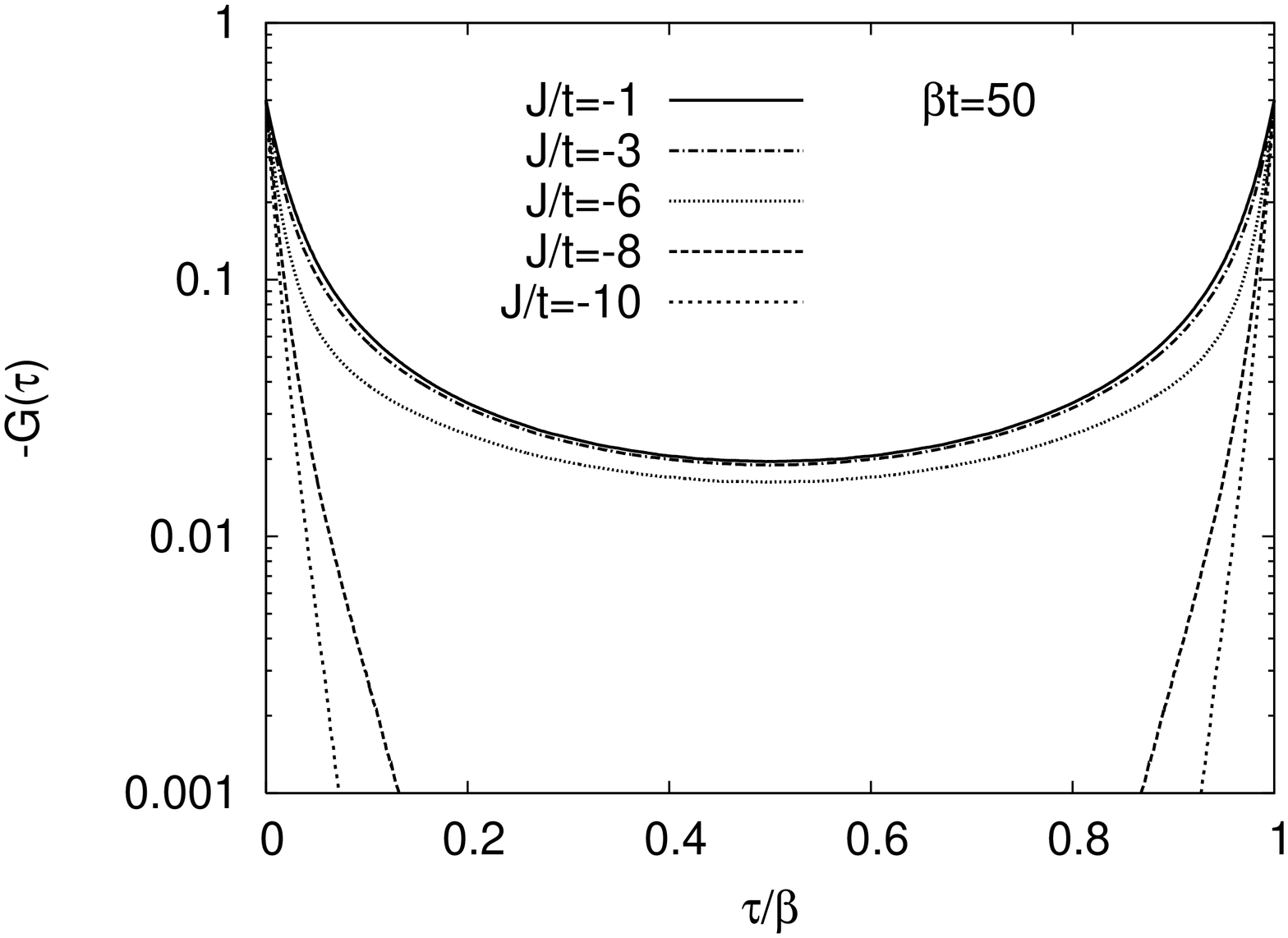}
\includegraphics [angle=-90, width=8.5cm] {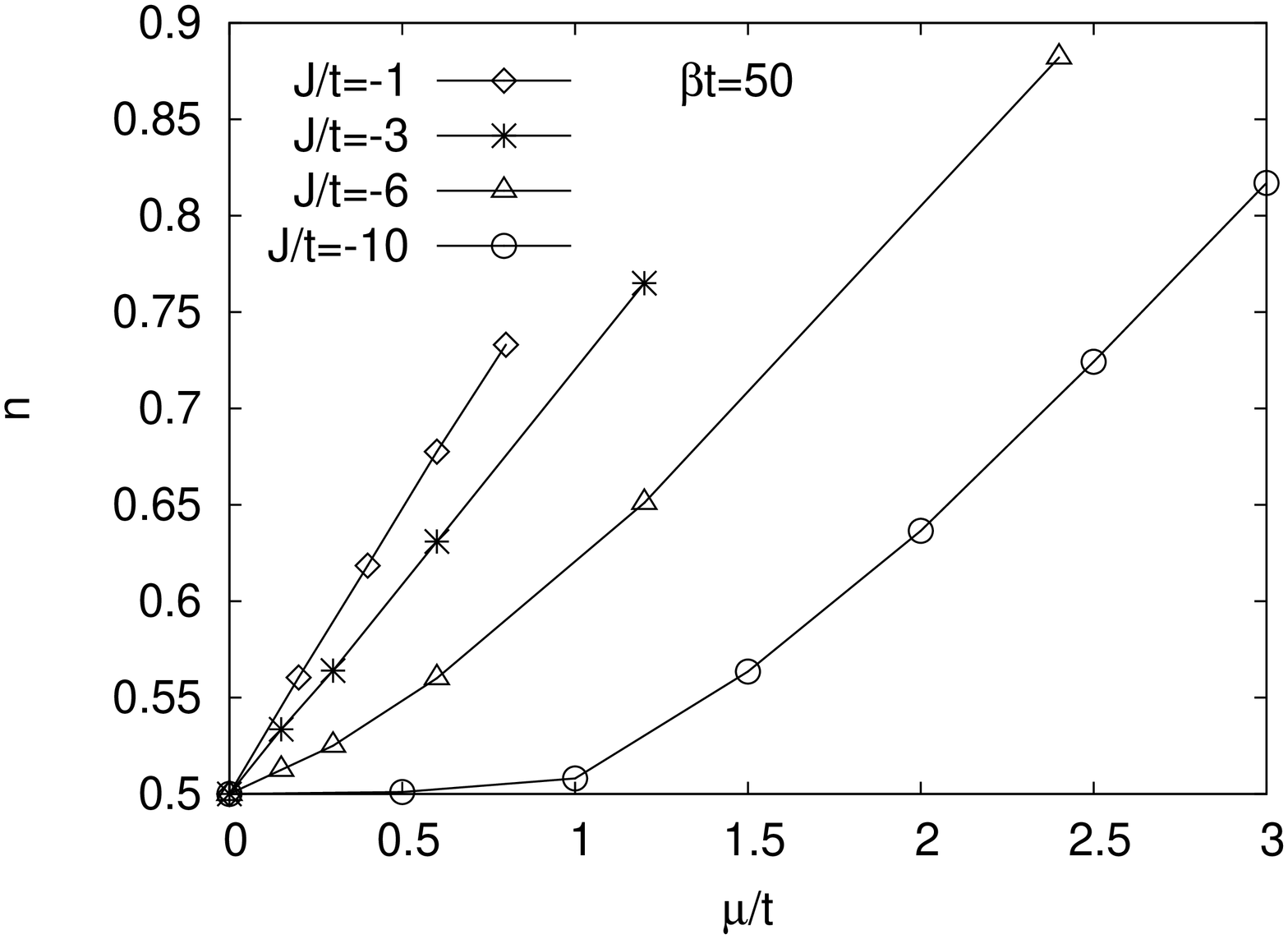}
\caption{Left panel: local Green functions for the half filled ferromagnetic Kondo lattice
model at the indicated exchange values and temperature. For $J/t\gtrsim -3$ the
computed Green functions are very close to the $J=0$ value and for $J/t\ge -6$ the long time behavior is characteristic of a metal. The exponential drop of the Green functions for $J/t\le -8$ shows that the system becomes insulating at these large couplings. Right panel: dependence of the density on chemical potential.  The smooth behavior for $J/t\ge-6$ shows the absence of a gap at half filling
whereas a gap is clearly evident in the curve corresponding to $J/t=-10$. Some suggestion of a precursor to a gap is visible at $J/t=-6$.}
\label{ferro}
\end{figure}

\begin{figure}[t]
\centering
\includegraphics [angle=-90, width=8.5cm] {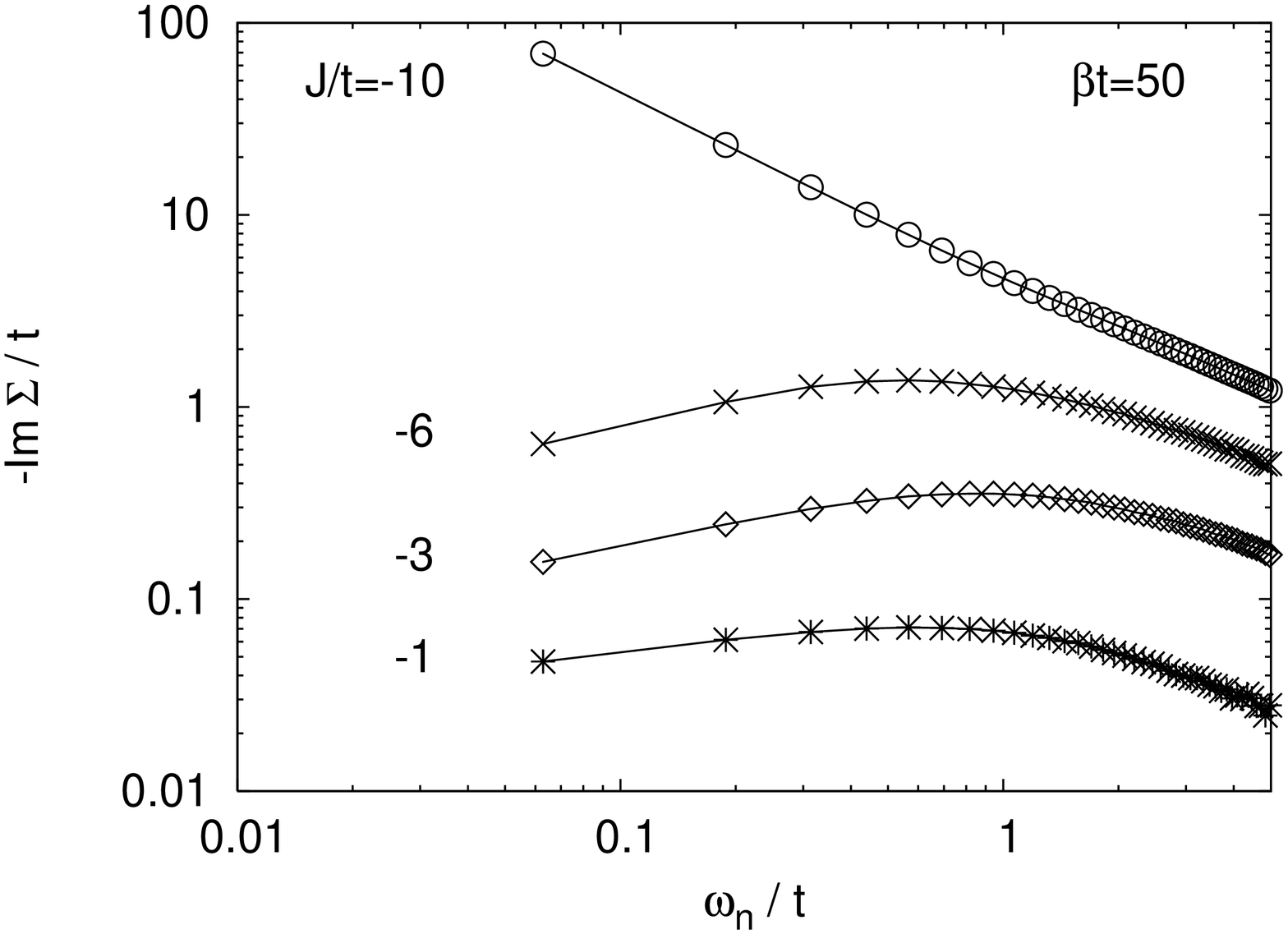}
\includegraphics [angle=-90, width=8.5cm] {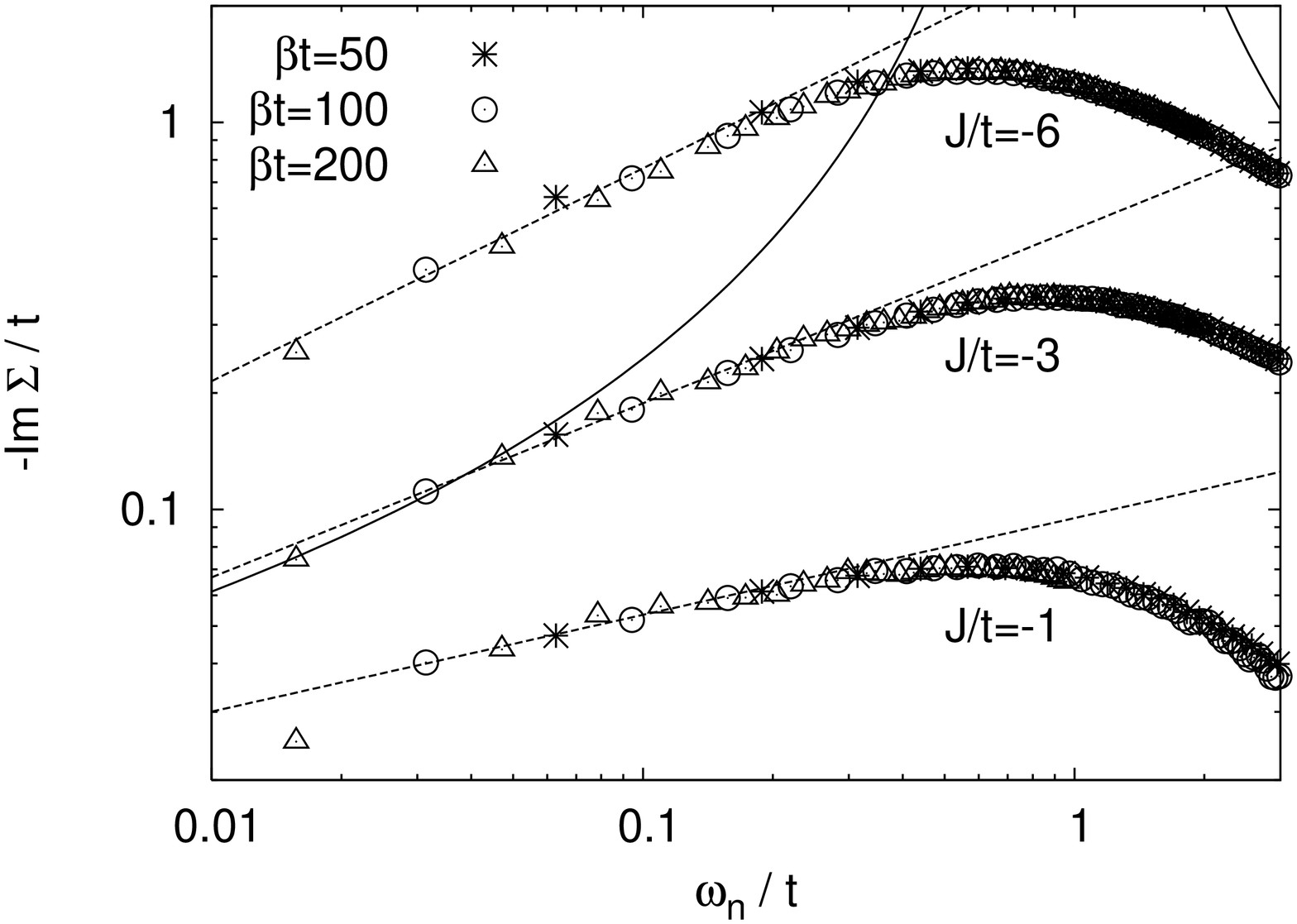}
\caption{
Left panel: imaginary part of the
electron self energy $\Sigma(i\omega_n)$ for the half-filled ferromagnetic Kondo lattice model at $\beta t=50$,  $J/t=-1$, $-3$, $-6$ and $-10$. The metal-insulator transition which takes place between the last two values of $J$ is obvious from the change in the low-frequency behavior.  Right panel: expanded view of the low frequency behavior of the self energy in the
smaller-$J$  ``metallic" phase. Dashed lines demonstrate an approximate power-law decrease of $\Im m \Sigma$ as $\omega \rightarrow 0$
with exponents $0.25$ for $J/t=-1$, $0.45$ for $J/t=-3$ and $0.55$ for $J/t=-6$. The solid line is proportional to the theoretically expected \cite{Biermann05} asymptotic behavior 
$\Im m \Sigma \sim 1/(\ln \omega_n)^2$.
}
\label{qp_log}
\end{figure}

We begin with the ferromagnetic case. 
The left hand panel of Fig.~\ref{ferro} plots 
the converged Green functions for the 
ferromagnetic couplings $J/t=-1$, $-3$, $-6$, $-8$ and $-10$ at half filling and at the low
temperature $\beta t=50$. 
It is apparent that for $J/t \gtrsim -6$ the Green function
is weakly dependent on $J$ and exhibits the slow decay with time 
characteristic of a metal.
As the exchange coupling is increased, the system eventually undergoes a metal-insulator 
transition at a critical value between $J/t=-6$ and $-8$ (see also Fig.~\ref{qp_log}), which is considerably
larger in magnitude than the classical-model critical value   $J/t=-4$ 
($J_\text{eff}=-t$).
The right panel shows the dependence of the particle number  per spin, $n$, 
on chemical potential for several $J$ values; for small coupling, a smooth evolution is seen with no 
indication of a gap, whereas the opening of a gap is evident in the $n(\mu)$-curve for $J/t=-10$.

\begin{figure}[t]
\centering
\includegraphics [angle=-90, width=8.5cm] {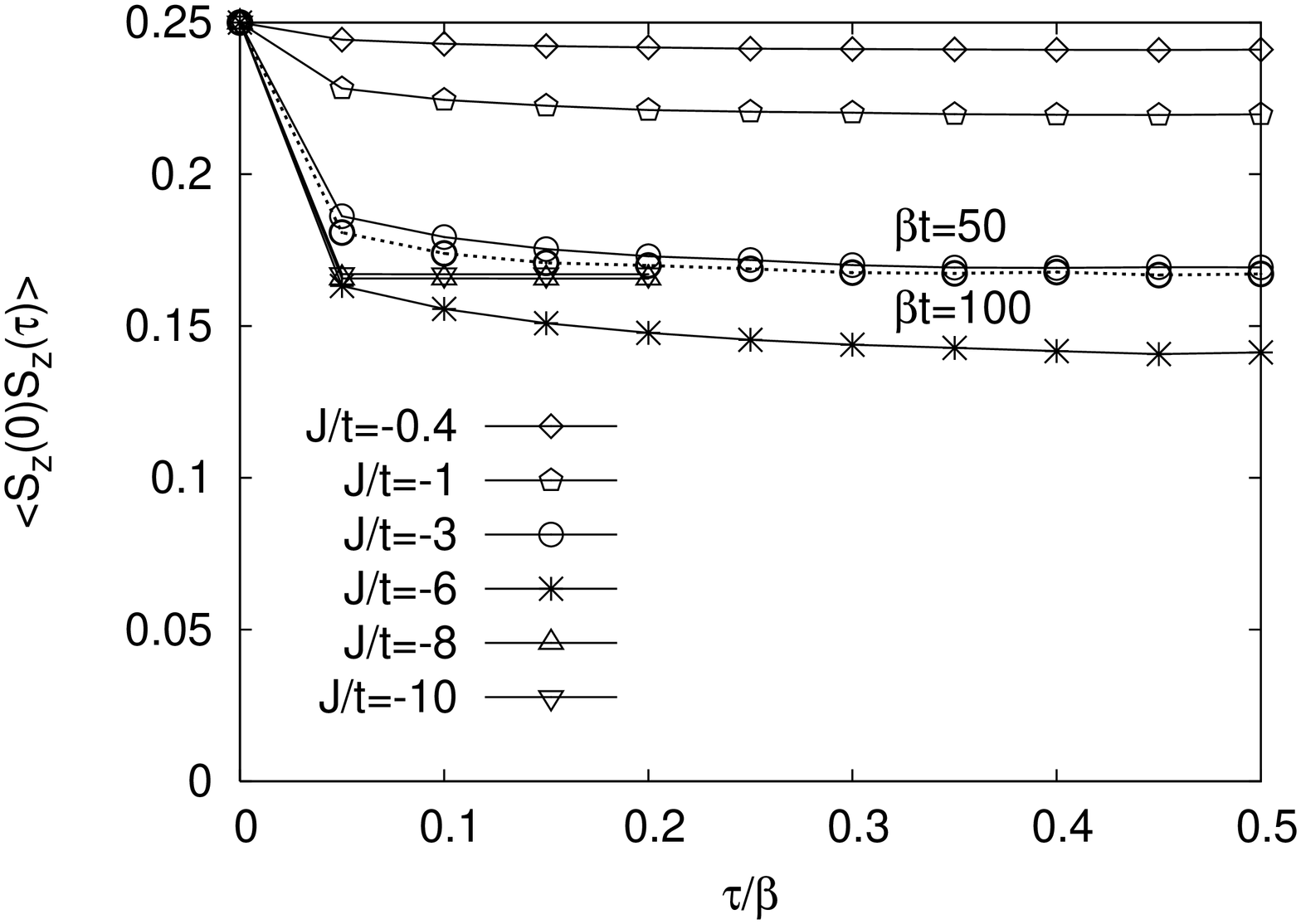}
\includegraphics [angle=-90, width=8.5cm] {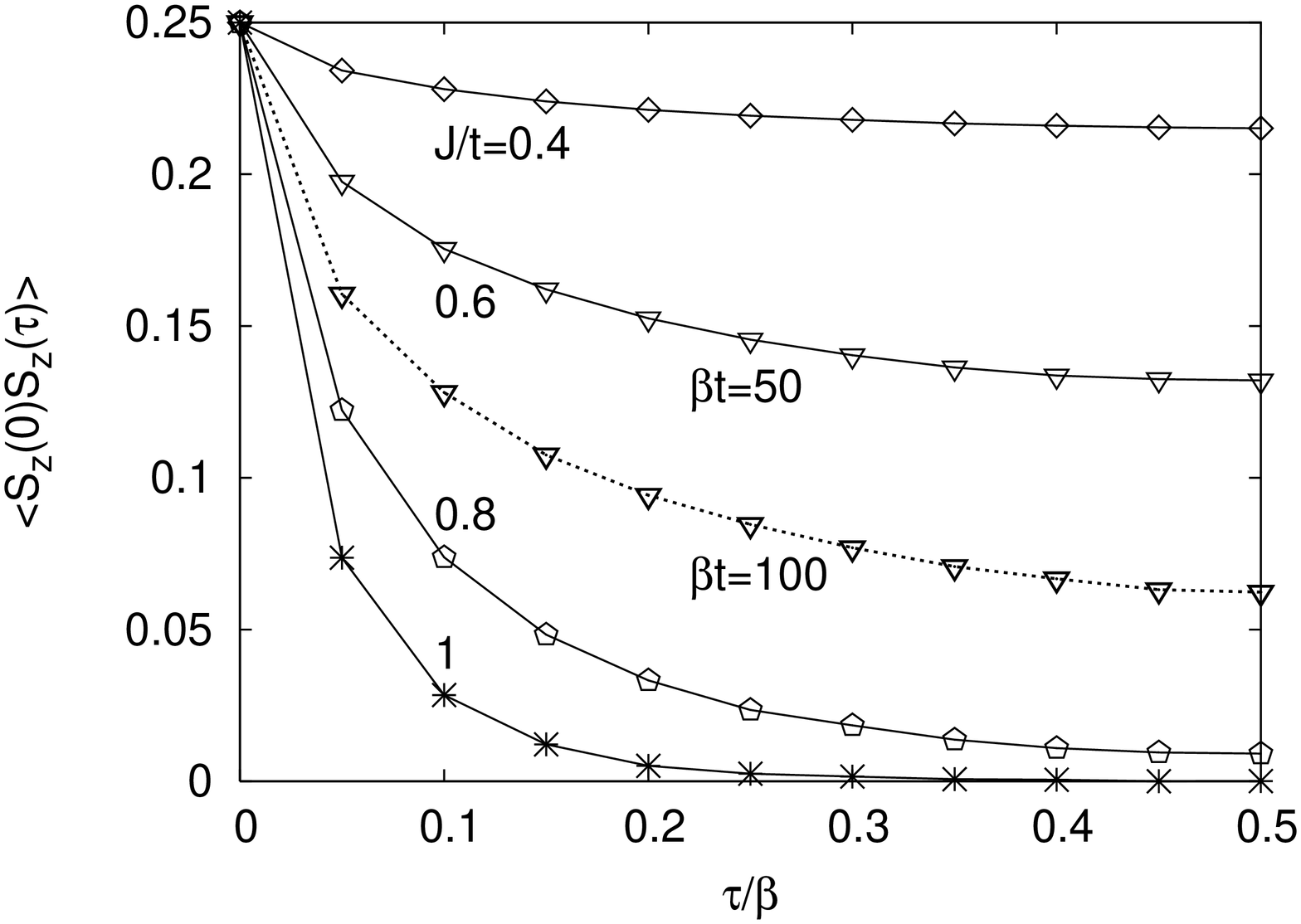}
\caption{
Imaginary time 
correlation function for the local moment calculated 
for ferromagnetic (left panel)
and antiferromagnetic (right panel) couplings  
at half filling for the $J$
values indicated. Solid lines show results for $\beta t=50$. 
The dotted lines
show results for  $\beta t=100$ and $J/t=-3$ (ferromagnetic case)
and $J/t=0.6$ (antiferromagnetic case).}
\label{corr}
\end{figure}

The left hand panel of Fig.~\ref{corr} shows the impurity-model spin-spin 
correlation function $C_{SS}(\tau)=\left<S_z(0)S_z(\tau)\right>$ calculated for the 
ferromagnetic Kondo lattice. An initial drop (with a $J$-dependent magnitude) is followed by a saturation
to an almost temperature-independent value. In the classical model in the paramagnetic phase
$C(\tau)=1/4$ independent of time. The   saturation seen in the quantum calculation thus indicates that
the long-time behavior of the spins is essentially classical, qualitatively consistent with the ferromagnetic
Kondo scaling discussed in Eq.~(\ref{RG}).  The combination of a paramagnetic state
and a saturated (non-vanishing) spin-spin correlator implies the  existence of annealed disorder
in the spins, in other words the existence of zero frequency spin fluctuations. In particular, the saturation 
evident in the  data for $J/t=-8$ and $-10$ shows that the charge gap seen in $G(\tau)$ does not imply
the opening of a spin gap.

In the classical model, the spin disorder in the paramagnetic phase leads to a non-vanishing self
energy at $\omega \rightarrow 0$ (either divergent, in the insulating phase, or finite,
in the metallic phase). Fig.~\ref{qp_log} shows the self energies calculated for the ferromagnetic quantum model. 
For $J/t=-10$, the system is insulating and $\Sigma$ diverges, as in the classical case. However, for the smaller $|J|$
the self energy clearly vanishes as $\omega \rightarrow 0$, a behavior quite different from that found in the classical case.

The differences between the quantum spin-ferromagnetic coupling calculations and the results for the classical
model have, we believe, a common origin, namely the decoupling of the carriers and spins at low energies (as found in the single-impurity model, Eq.~(\ref{RG})).
This is directly seen from the comparision of the spin-spin correlator
(which shows classical spins) and the metallic phase self energy (whose vanishing at small frequency suggests no
scattering at the fermi surface).  This physics was already noted by 
Biermann and co-workers \cite{Biermann05} in a study of a related model. These authors argued
that one could obtain the low frequency behavior of the electron self energy by combining the 
Kondo scaling Eq.~(\ref{RG}) with the perturbative formula for the self energy to obtain
$\Sigma(\omega)\approx \rho J_\text{eff}^2(\omega) \sim (1+\ln \rho \omega)^{-2}$.
The right hand panel of Fig.~\ref{qp_log} shows an expanded view of the lower frequency regime of the metallic phase self energies.
One sees that at the frequencies accessible to us the self energy is better fitted by a weak, $J$-dependent power law 
(the dashed lines correspond to the exponents $0.25$ for $J/t=-1$, $0.45$ for $J/t=-3$ and $0.55$ for $J/t=-6$). 
In particular, except perhaps at $J/t=-1$, the curvature of the numerical data is opposite to the curvature predicted
by the one-impurity form. We suggest that the power law arises from an interplay between the 
one-impurity ferromagnetic Kondo scaling and the density of states renormalization due to $J$.
In particular, for $J$ near the critical value for the metal-insulator transition one expects a vanishing density
of states. However, we note that the temperature range is insufficient to rule out a low-$T$ crossover to the form proposed
in Ref.~\cite{Biermann05}. Further study of the frequency dependence of the self energy, and  in particular
a more precise characterization of the power law 
associated with the metal-insulator
critical point, would be of great interest.  

\subsubsection{Antiferromagnetic $J$}

\begin{figure}[t]
\centering
\includegraphics [angle=-90, width=8.5cm] {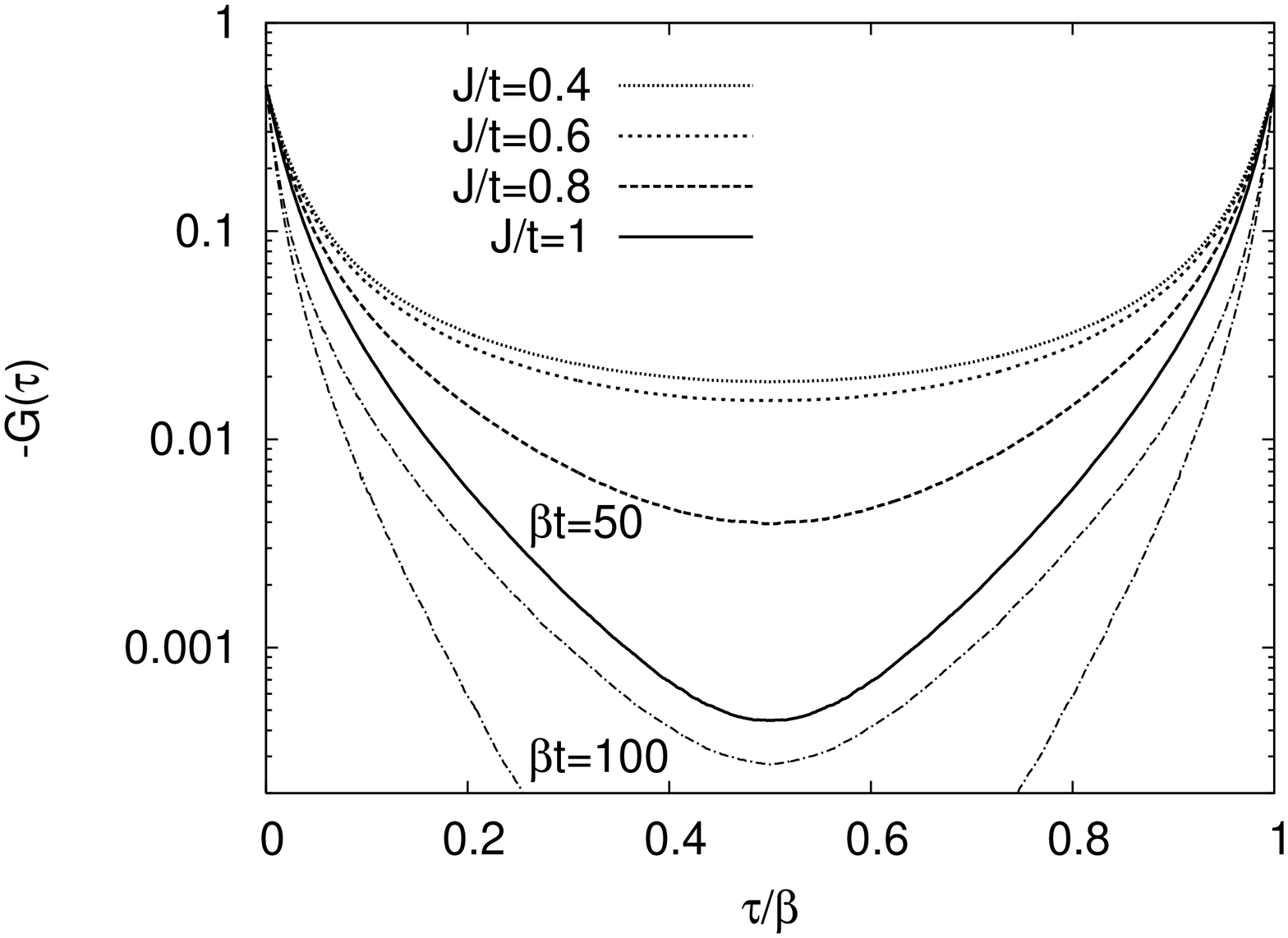}
\includegraphics [angle=-90, width=8.5cm] {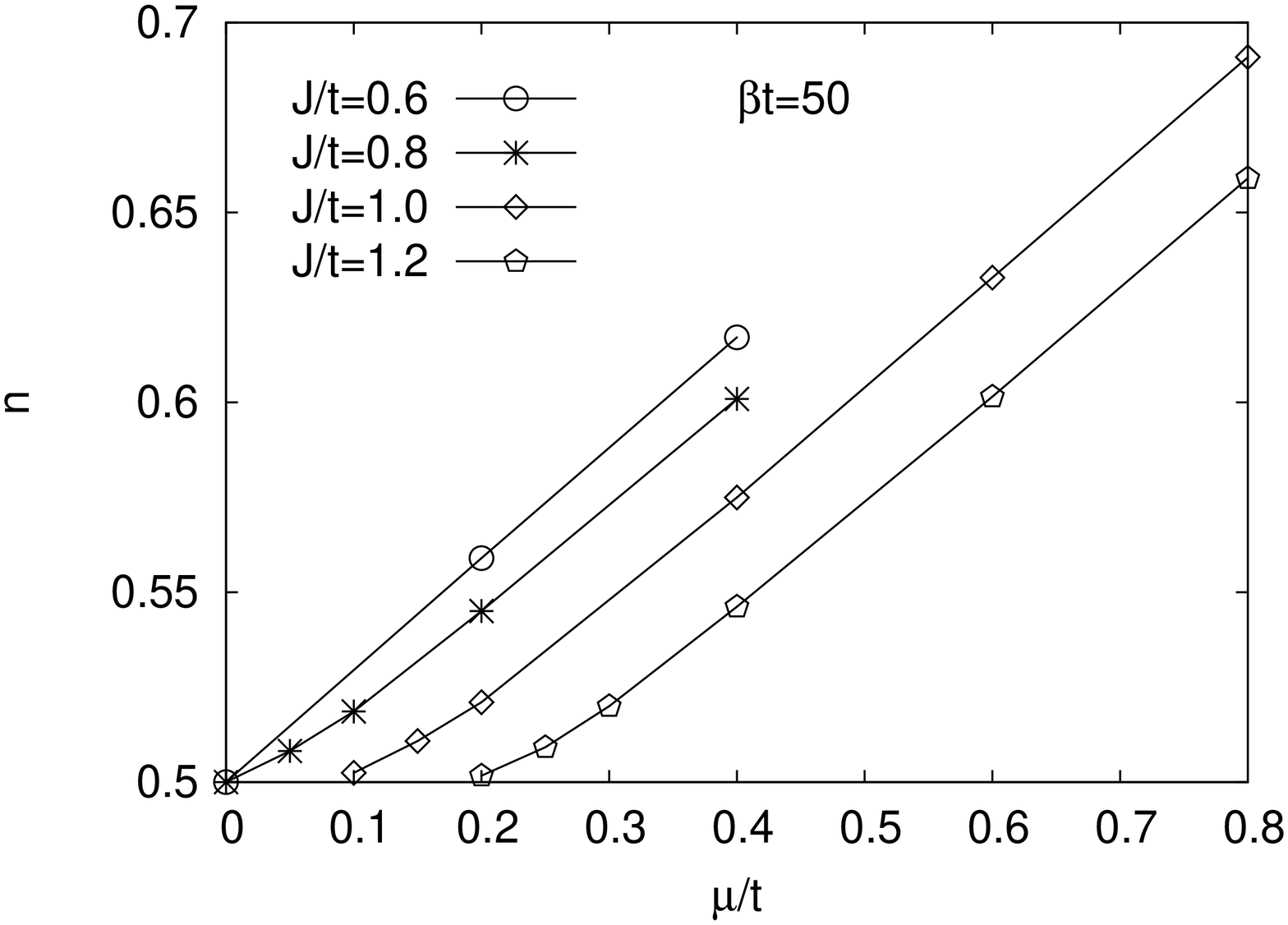}
\caption{Left panel: thick lines show the local Green functions for the antiferromagnetic Kondo lattice model for
$J/t=0.4,0.6,0.8,1$
and inverse temperature $\beta t=50$. Thin dash-dotted lines correspond to $\beta t=100$ and $J/t=0.8,1 $.
Right panel: density per spin
plotted as a function of chemical potential. The data for $J/t=1.0$  and 1.2 are consistent
with the opening of a charge gap. }
\label{antiferro}
\end{figure}

The physics of the antiferromagnetic Kondo lattice is markedly different from that of
either the classical spin or the ferromagnetic $S=1/2$ Kondo lattice. The left panel of 
Fig.~\ref{antiferro} shows the electron Green function calculated for several (small) $J$
values and low temperatures. Comparison to Fig.~\ref{ferro} shows that for all $J/t>0.4$,
$G(\beta/2)$ falls below the value $4/(\pi \beta t) \approx 0.0254$ expected for a fermi liquid
and approximately observed in the ferromagnetic case.  Furthermore,  as $T$ is decreased $G$ drops
rapidly, suggesting the opening of the gap expected for a Kondo insulator \cite{kondo_gap}. 
We believe that even the smallest $J$ will eventually become insulating,
but that the gap is too small to be seen on the temperature scales we have studied.   The right hand
panel shows that for $J/t=1.0$ and $1.2$ a gap in the excitation spectrum is evident in the 
$n(\mu)$ curve. Also, as expected in the presence of a charge gap, we find that the 
imaginary part of the self-energy diverges as $\omega_n\rightarrow 0$ (not shown).

\begin{figure}[t]
\centering
\includegraphics [angle=-90, width=8.5cm] {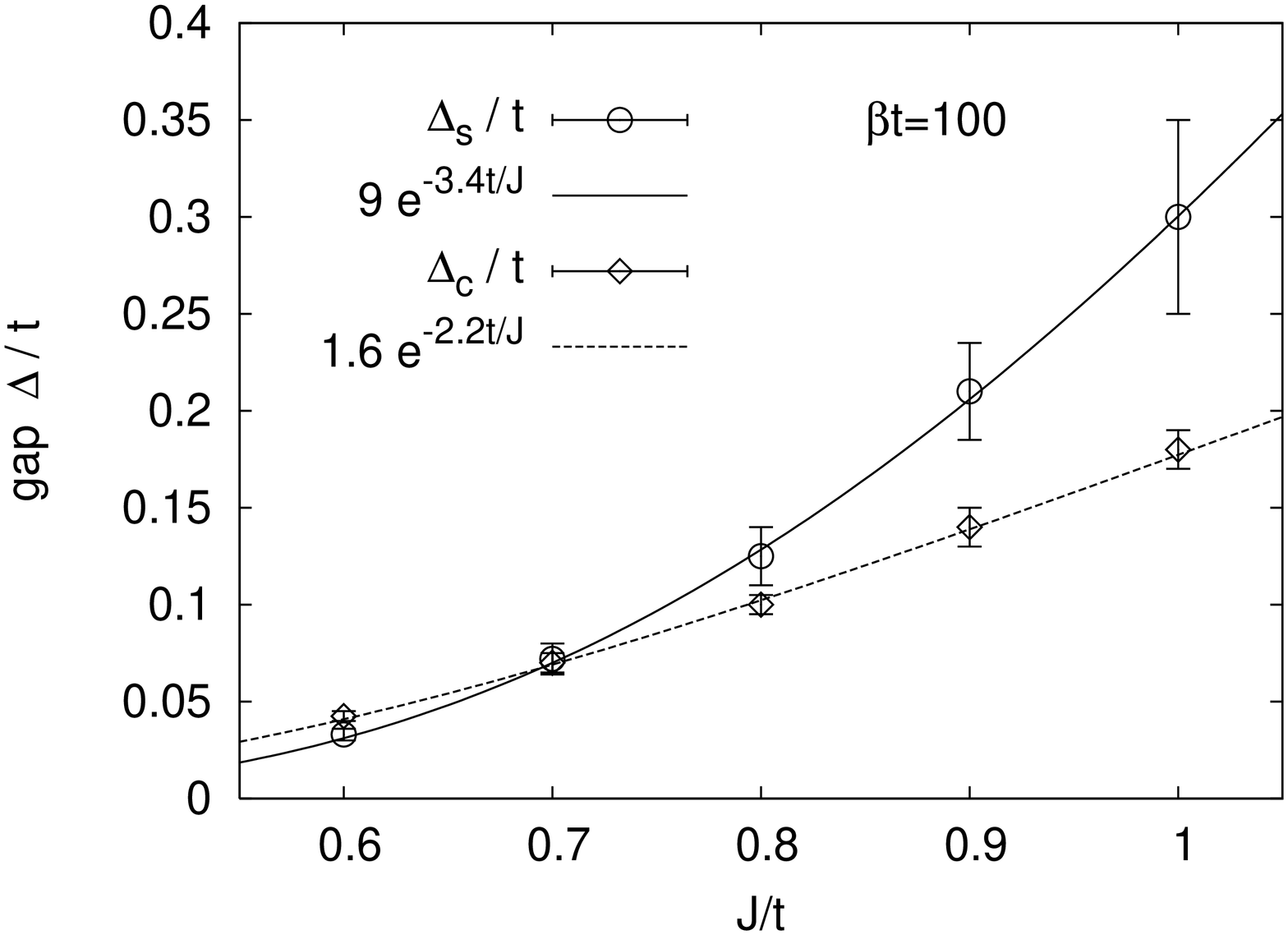}
\includegraphics [angle=-90, width=8.5cm] {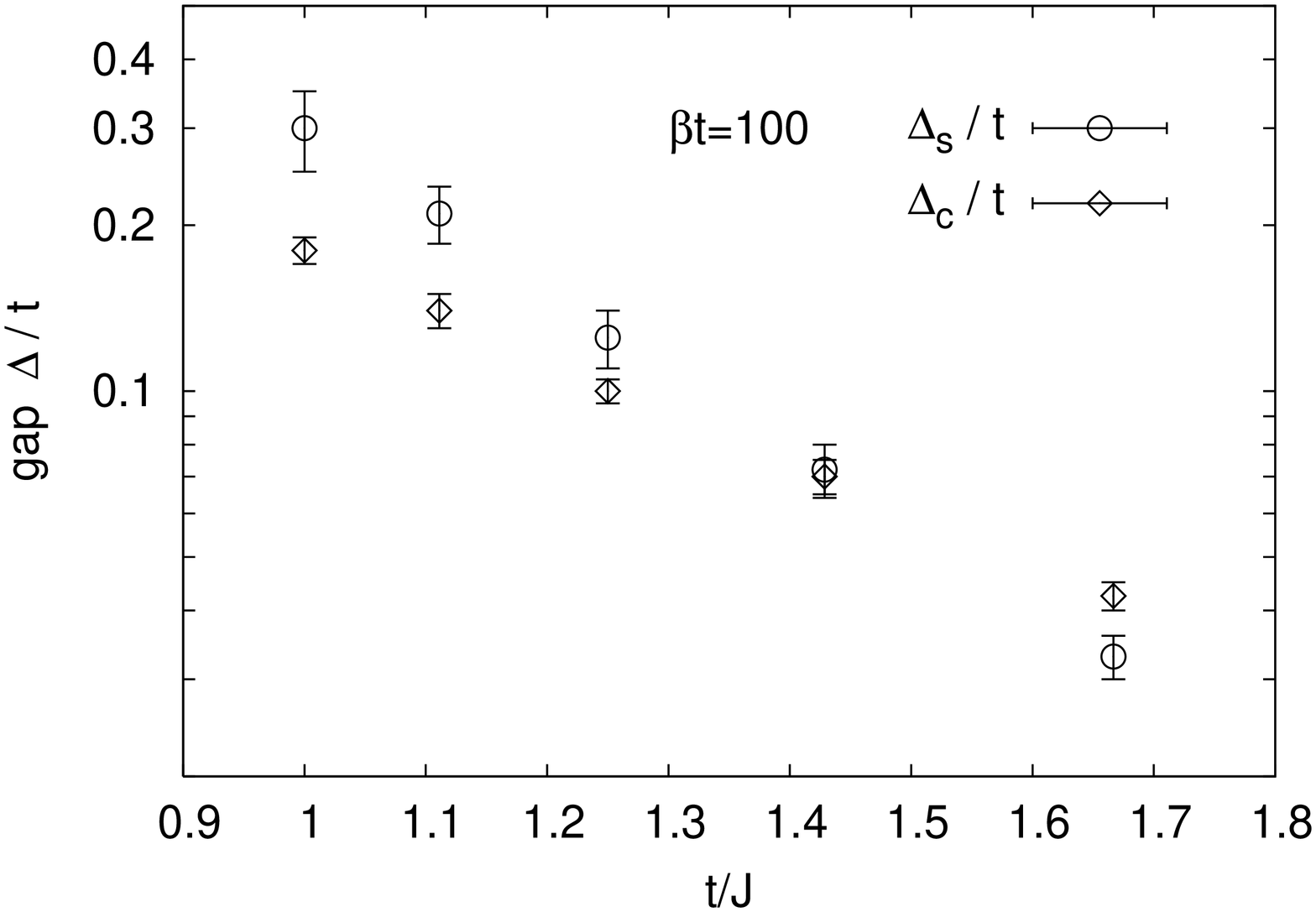}
\caption{
Left panel: spin gap $\Delta_s$ and charge gap $\Delta_c$ extracted from fits of the function $a\cosh(\Delta(\tau-\beta/2))$ to the 
spin-spin correlation functions and Green functions obtained for $\beta t= 100$ and the indicated values of $J$. Right panel: same data plotted as a function of $t/J$ on a semi-$\log$ scale. The results are consistent with the expected small-$J$ behavior $\ln \Delta\sim -1/\rho J$. 
}
\label{gap}
\end{figure}

The spin spin correlation functions for antiferromagnetic coupling are shown in the right hand panel
of Fig.~\ref{corr}.  The correlations decay rapidly with time, consistent with the formation of a gapped Kondo insulating 
state.  While the exponential decay may not be obvious from the $\beta t=50$ data, their $\beta t=100$ counterparts 
(shown as an illustration for $J/t=0.6$ by the dotted line) can be reasonably well fitted to a function of the form 
$a\cosh(\Delta_s(\tau-\beta/2))$. 
From these fits we extract the spin gaps $\Delta_s$ shown in Fig.~\ref{gap}. Also plotted are the charge gaps $\Delta_c$, which we obtained from an analogous fit to the Green functions.
The variation of the gaps with $J$ is very rapid 
and (as shown in the right hand panel of Fig.~\ref{gap}) is roughly consistent with  the theoretically expected behavior $\ln \Delta \sim -1/\rho J$ at small $J$, crossing over to $\Delta \sim J$ for $J>t$. Remarkably, we find that the impurity model spin gap is less than twice the charge gap, with the ratio $\Delta_s/\Delta_c$ decreasing through 1 as $J$ is decreased. We understand this as a precursor of the magnetic state which would exist at small $J$ and low $T$ if magnetic order were not suppressed. 
However, we caution the reader that the spin gaps at the larger $J$-values are so large they are difficult to determine accurately, while the charge gap is uncertain at small $J$ because the Green functions do not very nicely fit to a $\cosh$-function.

\subsection{Magnetic Ordering}

We now show that our method correctly captures the magnetic ordering phenomena characteristic of the Kondo lattice.
As in the previous sections, we specialize to half filling, bipartite lattices, and particle-hole symmetry.  
For orientation, we first review the known results for the classical-spin  case.
At half filling the classical model has antiferromagnetic order at all coupling strengths \cite{Chattopadhyay01}.  
At very small $J$, the classical transition temperature grows as $T^\text{cl}_N \sim J^2/t$. It
reaches a maximum around $J/t\approx 2$ and for large $J$ decreases as $T^\text{cl}_N\sim t^2/J$. 
In the quantum ferromagnetic case, we expect the $T^\text{cl}_N(J)$-curve to retain essentially the same shape. 
In the quantum antiferromagnetic case we expect a quantum phase transition to a singlet phase for $J$ larger than a critical value
\cite{Doniach77,kondo_gap}.

\begin{figure}[t]
\centering
\includegraphics [angle=-90, width=8.5cm] {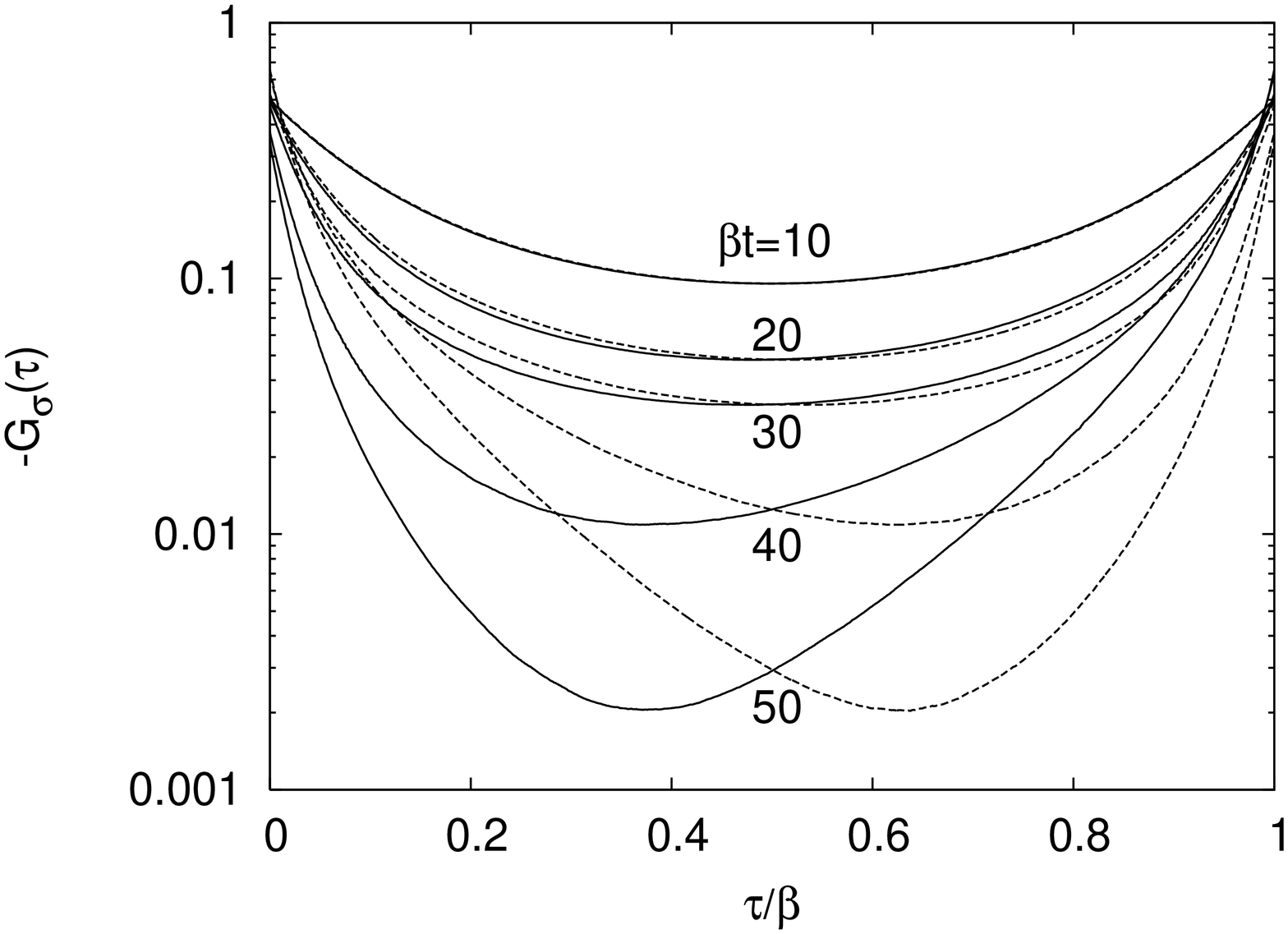}
\includegraphics [angle=-90, width=8.5cm] {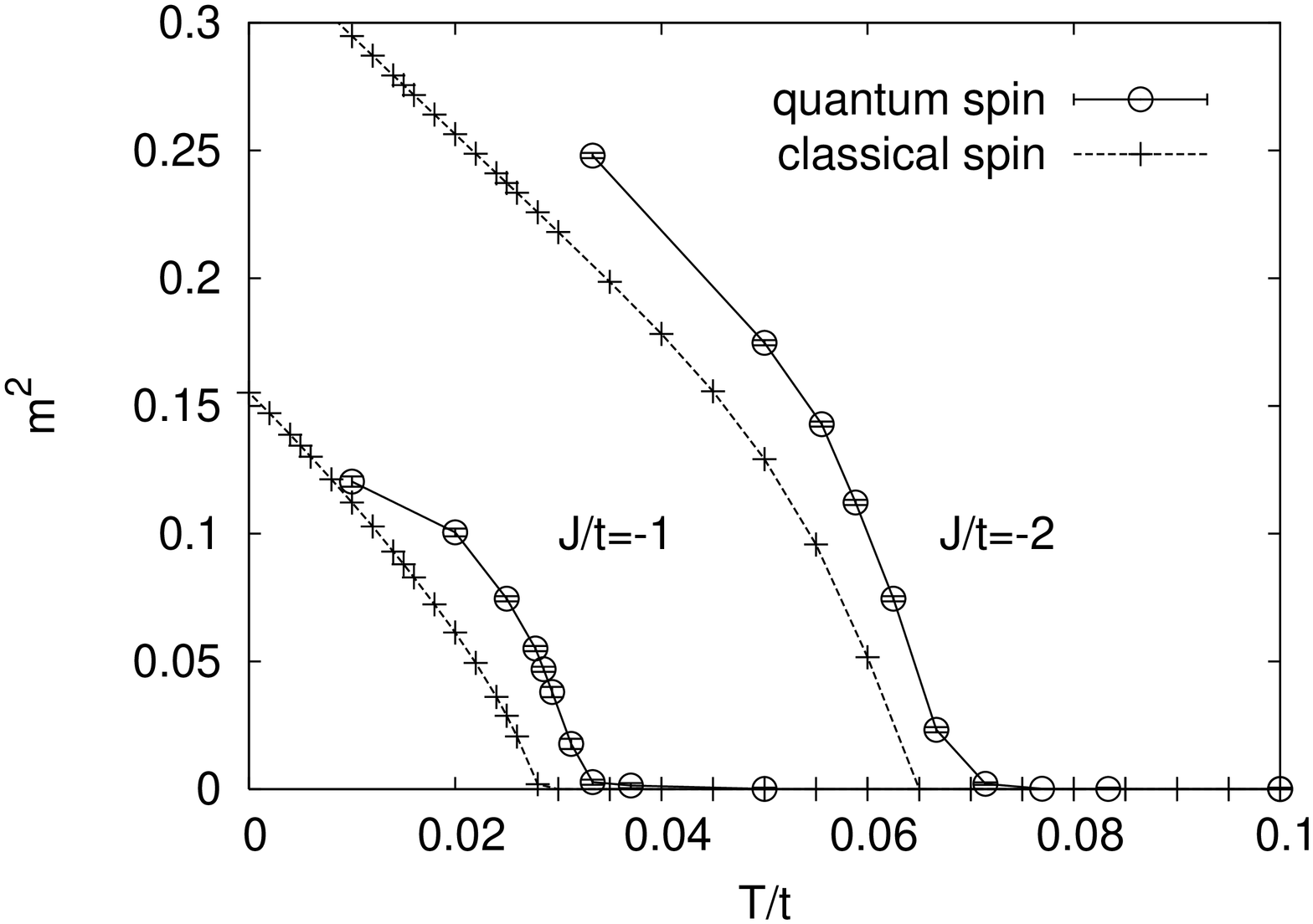}
\caption{
Ferromagnetically coupled kondo lattice. Left panel: Green functions obtained for $J/t=-1$ 
and $\beta t=10$, 20, 30, 40 and 50. A magnetic transition, setting in at  
$\beta t\approx 30$ is evident from the appearance of a difference between spin up and spin down
Green functions. Right panel: staggered magnetization $m=n_\uparrow-n_\downarrow$ as a function of  temperature. 
Solid lines: $S=1/2$ model, $J/t=-1$ and $-2$. Dashed lines: results from the classical model
for $J$ corresponding to the same diagonal spin splitting. 
}
\label{magnetic_ferro}
\end{figure}

We now turn to the results for the quantum model, beginning with ferromagnetic couplings.
At half filling ferromagnetism is never found to be stabilized, whereas the left panel of 
Fig.~\ref{magnetic_ferro}  shows that with use of the antiferromagnetic self consistency condition 
a spin polarization (difference between up and down Green functions) becomes apparent for
$\beta t \gtrsim 20$ and $J/t=-1$. The spin polarization is associated with the formation of a gap, as can be seen from the rapid time decay of the lower-$T$ Green functions (compared for example to the paramagnetic solution for $\beta t=50$ in Fig.~\ref{ferro}). Hence the ground state of the ferromagnetic Kondo lattice model is an antiferromagnetically ordered insulator.

The right-hand panel of Fig.~\ref{magnetic_ferro}  shows the staggered magnetization. Around $T/t = 0.033$, the staggered magnetization $m=n_\uparrow-n_\downarrow$ for $J/t=-1$ starts to increase rapidly. 
We also plot data for a larger magnitude coupling 
$J/t=-2$, as well as results calculated for the classical model at couplings corresponding to the same effective
spin splitting (dashed lines) \cite{Lin06}.  
Surprisingly, in view of the ferromagnetic Kondo scaling, the critical temperatures in both models are comparable. While the magnetization onset in the quantum spin case is more rapid, the low-$T$ saturation value is apparently lower. 

\begin{figure}[t]
\centering
\includegraphics [angle=-90, width=8.5cm] {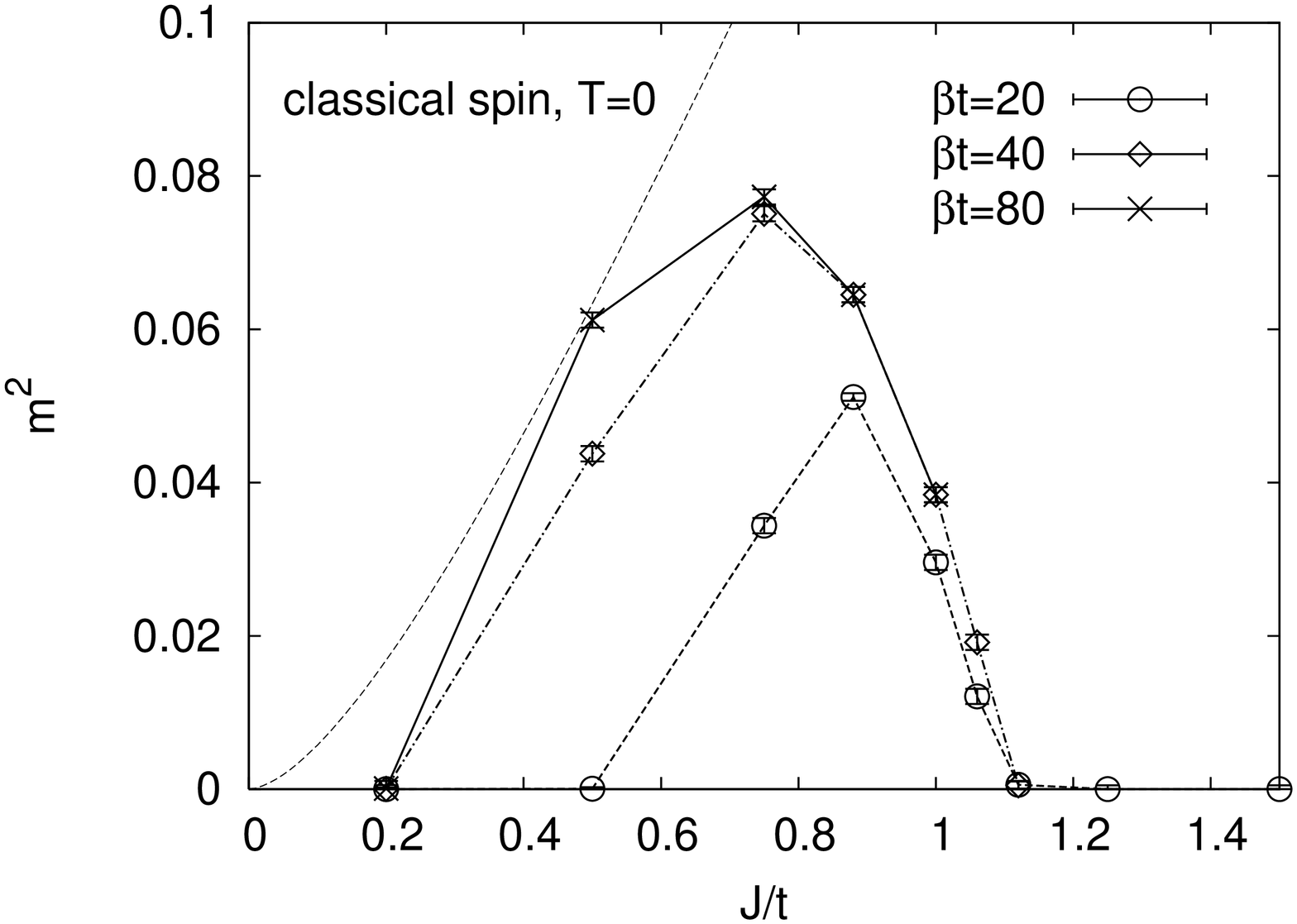}
\includegraphics [angle=-90, width=8.5cm] {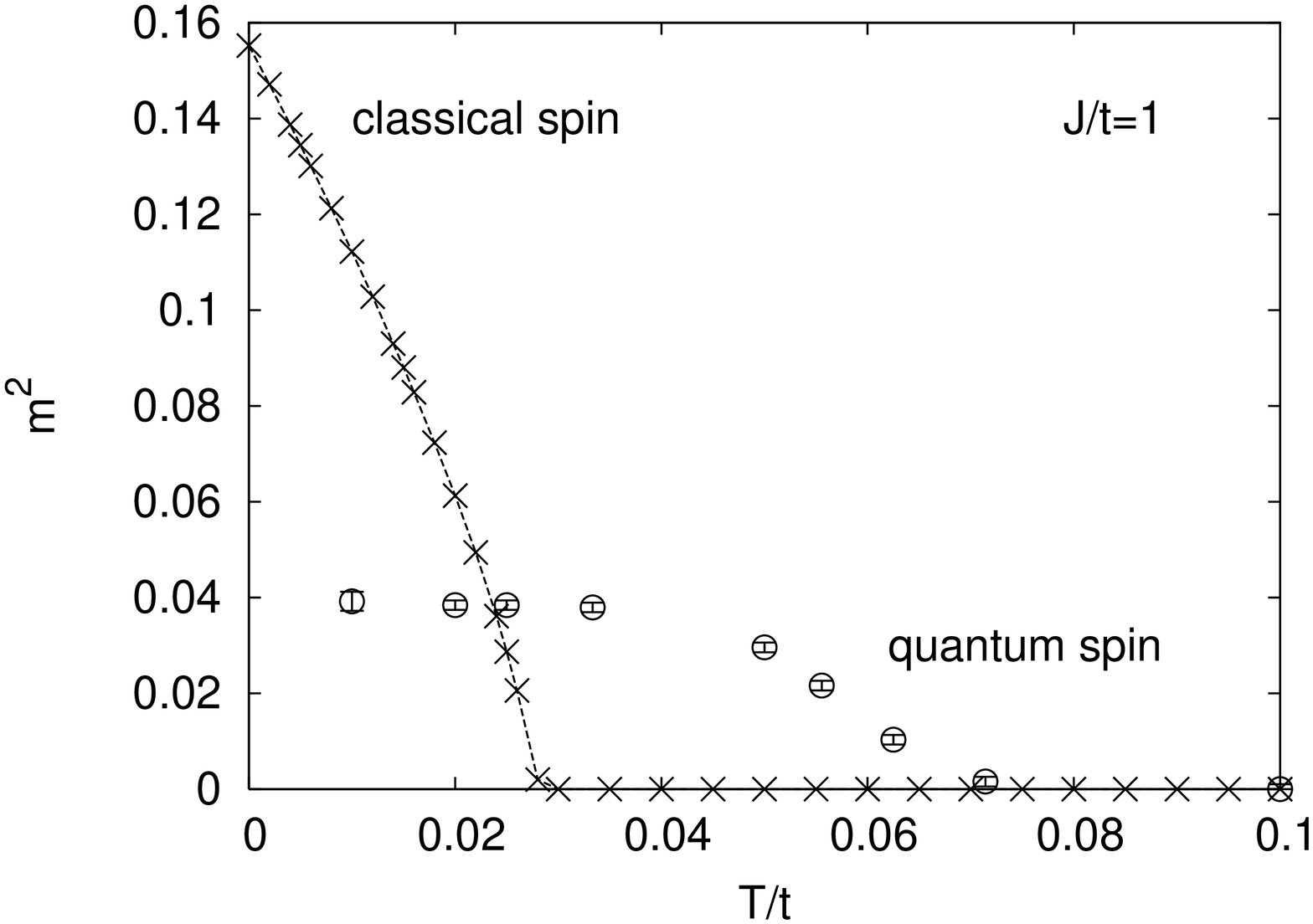}
\caption{
Staggered magnetization of the antiferromagnetically coupled Kondo lattice  model (half filling, bipartite lattice, particle-hole symmetry). 
Left panel: staggered magnetization as a function of $J/t$ for $\beta t = 20$, $40$ and $80$. 
There is an antiferromagnetic state at small coupling (for sufficiently low temperature) and around $J/t=1$ a quantum phase transition to a paramagnetic insulator. The dashed line shows the $T=0$ result for classical spins. The right panel plots $m=n_\uparrow-n_\downarrow$ as a function of temperature. We find that the transition temperature is considerably higher than for ferromagnetic coupling and that the magnetization saturates at a smaller value. This smaller magnetization is the result of stronger quantum fluctuations (Kondo divergence) and singlet formation. 
}
\label{magnetic_antiferro}
\end{figure}

On the mean field level, one expects a continuous transition of the form $m^2\sim T_c-T$. In the quantum case, we find $m^2(T)$-curves which are roughly consistent with the linear behavior of the classical model, although the magnetization drops somewhat more rapidly near the critical temperature.
The numerical data for $J/t=-1$ might even hint a first order transition. In addition to the steep drop near $T_c$, an essentially paramagnetic solution remains apparently stable for some range of temperatures below $T_c$. However, a definite statement would require a more detailed investigation of the behavior near the critical point. 

In Fig.~\ref{magnetic_antiferro} we show the staggered magnetization of   the antiferromagnetically coupled model as a function
of $J/t$ at several fixed temperatures. On the small $J$ side a strong temperature dependence is evident, reflecting the
strong $J$ dependence of the Neel temperature at weak coupling. For $J/t\gtrsim 0.75$ the $\beta t=40$ data provide a good estimate of the $T=0$ result. At $J/t \gtrsim 1$ the staggered magnetization rapidly drops to zero. This is the quantum phase transition to the singlet, Kondo insulator phase.
We observe
that this phase transition occurs at a $J$ which is small relative to the bandwidth. The dashed line indicates the $T=0$ result for classical spins. In this case, no transition to a paramagnetic insulator occurs. 
The right hand panel again shows the magnetization as a function of temperature. For $J/t=1$, magnetic order sets in around $T/t = 0.077$, which is  noticeably higher than the transition temperature
of the ferromagnetically coupled model or the model with classical spins. We attribute this to the growth in $J$ implied by the antiferromagnetic
Kondo scaling.  On the other hand, due to the  tendency to form singlets, the magnetization $m=n_\uparrow -n_\downarrow$ for the antiferromagnetic model with quantum spins saturates at $m\approx 0.2$, 
which is considerably smaller than the staggered magnetization  of the corresponding ferromagnetic system.

\section{Two orbital model}

For a second demonstration of the power of the method we consider here the two orbital model studied by other workers as a model for the orbital selective Mott transition.
The local Hamiltonian is 
\begin{eqnarray}
H_\text{loc}&=&-\sum_{\alpha=1,2}\sum_{\sigma}\mu n_{\alpha,\sigma} + \sum_{\alpha=1,2} U n_{\alpha,\uparrow} n_{\alpha,\downarrow} 
+ \sum_\sigma U' n_{1,\sigma} n_{2,-\sigma} + \sum_\sigma (U'-J) n_{1,\sigma}n_{2,\sigma}\nonumber\\
&&-J(\psi^\dagger_{1,\downarrow}\psi^\dagger_{2,\uparrow}\psi_{2,\downarrow}\psi_{1,\uparrow}
+ \psi^\dagger_{2,\uparrow}\psi^\dagger_{2,\downarrow}\psi_{1,\uparrow}\psi_{1,\downarrow} + h.c.).
\end{eqnarray}
We adopt the conventional choice of parameters, $U'=U-2J$,
which follows from symmetry considerations for $d$-orbitals in free space and is also assumed to hold in solids. We consider semi-circular densities of states of bandwith $4t_1$ and $4t_2$ for orbitals 1 and 2, respectively, with a fixed ratio $t_2/t_1=2$, and furthermore restrict ourselves to the paramagnetic phase by averaging over spin in each orbital.  

The half-filling condition for this model is $\mu=\frac{3}{2}U-\frac{5}{2}J$ and the 16 eigenstates and their energies are listed in Tab.~\ref{eigenstates_2orbital}.
\begin{table}[t]
\begin{tabular}{llll}
Eigenstates \hspace{20mm}\mbox& Energy\hspace{30mm} &Eigenstates \hspace{20mm}\mbox& Energy\\
&\\
$|1\rangle = | 0,0\rangle$ & 0 \vspace{1mm}&
$|9\rangle = \frac{1}{\sqrt{2}}(|\!\uparrow,\downarrow\rangle-|\!\downarrow, \uparrow \rangle)$ & $U-J-2\mu$ \vspace{1mm}\\
$|2\rangle = |\!\uparrow,0\rangle$ & $-\mu$ \vspace{1mm}&
$|10\rangle = \frac{1}{\sqrt{2}}(|\!\uparrow\downarrow,0\rangle-|0, \uparrow\downarrow \rangle)$\hspace{5mm} & $U-J-2\mu$ \vspace{1mm}\\
$|3\rangle = |\!\downarrow,0\rangle$ & $-\mu$ \vspace{1mm}&
$|11\rangle =  \frac{1}{\sqrt{2}}(|\!\uparrow\downarrow,0\rangle+|0, \uparrow\downarrow \rangle)$ & $U+J-2\mu$ \vspace{1mm}\\
$|4\rangle = | 0,\uparrow\rangle$ & $-\mu$ \vspace{1mm}&
$|12\rangle = | \!\uparrow\downarrow,\uparrow\rangle$ & $3U-5J-3\mu$ \vspace{1mm}\\
$|5\rangle = | 0,\downarrow\rangle$ & $-\mu$ \vspace{1mm}&
$|13\rangle = | \!\uparrow\downarrow,\downarrow\rangle$ & $3U-5J-3\mu$ \vspace{1mm}\\
$|6\rangle = |\!\uparrow,\uparrow\rangle$ & $U-3J-2\mu$ \vspace{1mm}&
$|14\rangle = |\!\uparrow,\uparrow\downarrow\rangle$ & $3U-5J-3\mu$ \vspace{1mm}\\
$|7\rangle = \frac{1}{\sqrt{2}}(|\!\uparrow,\downarrow\rangle+|\!\downarrow, \uparrow \rangle)$\hspace{5mm} & $U-3J-2\mu$ \vspace{1mm}&
$|15\rangle = |\!\downarrow,\uparrow\downarrow\rangle$ & $3U-5J-3\mu$ \vspace{1mm}\\
$|8\rangle = |\!\downarrow,\downarrow\rangle$ & $U-3J-2\mu$ \vspace{1mm}&
$|16\rangle = |\!\uparrow\downarrow,\uparrow\downarrow\rangle$ & $6U-10J-4\mu$ \vspace{1mm}\\
\end{tabular}
\caption{Eigenstates and eigenenergies for the local part of the 2-orbital model. The first entry corresponds to orbital 1 and the second entry to orbital 2.}
\label{eigenstates_2orbital}
\end{table}
In this basis, the propagators $K$ are diagonal, while the creation and annihilation operators for the different orbital and spin states become sparse $16 \times 16$ matrices. For a given spin, no more than two creation (or annihilation) operators may occur in a row and we check this condition before actually computing the trace.

An issue of debate in recent years has been the occurrence of an orbital selective Mott transition in such 2-orbital systems with Hund's coupling and different band widths, $t_1\ne t_2$. Using exact diagonalization \cite{Caffarel94} to solve the impurity problem, Koga {\it et al.} \cite{Koga04} found that the narrow band becomes insulating at a smaller coupling than the wide band. For semi-circular densities of states with a ratio of band widths $t_2/t_1=2$, the critical couplings were found to be approximately $U^c_1/t_1=5.4$ and $U^c_2/t_1=7$.  On the other hand, in earlier work using QMC simulations, Liebsch \cite{Liebsch03} concluded that the transition takes place simultaneously in both bands. The QMC method should be more reliable than a ED calculation with a small number of bath sites, but the straight forward extension of the usual auxiliary field approach \cite{Hirsch86} suffers from a bad sign problem in the presence of spin flip and pair hopping processes, which were thus ignored in Ref.~\cite{Liebsch03}. 
\begin{figure}[t]
\centering
\includegraphics [angle=-90, width=8.5cm] {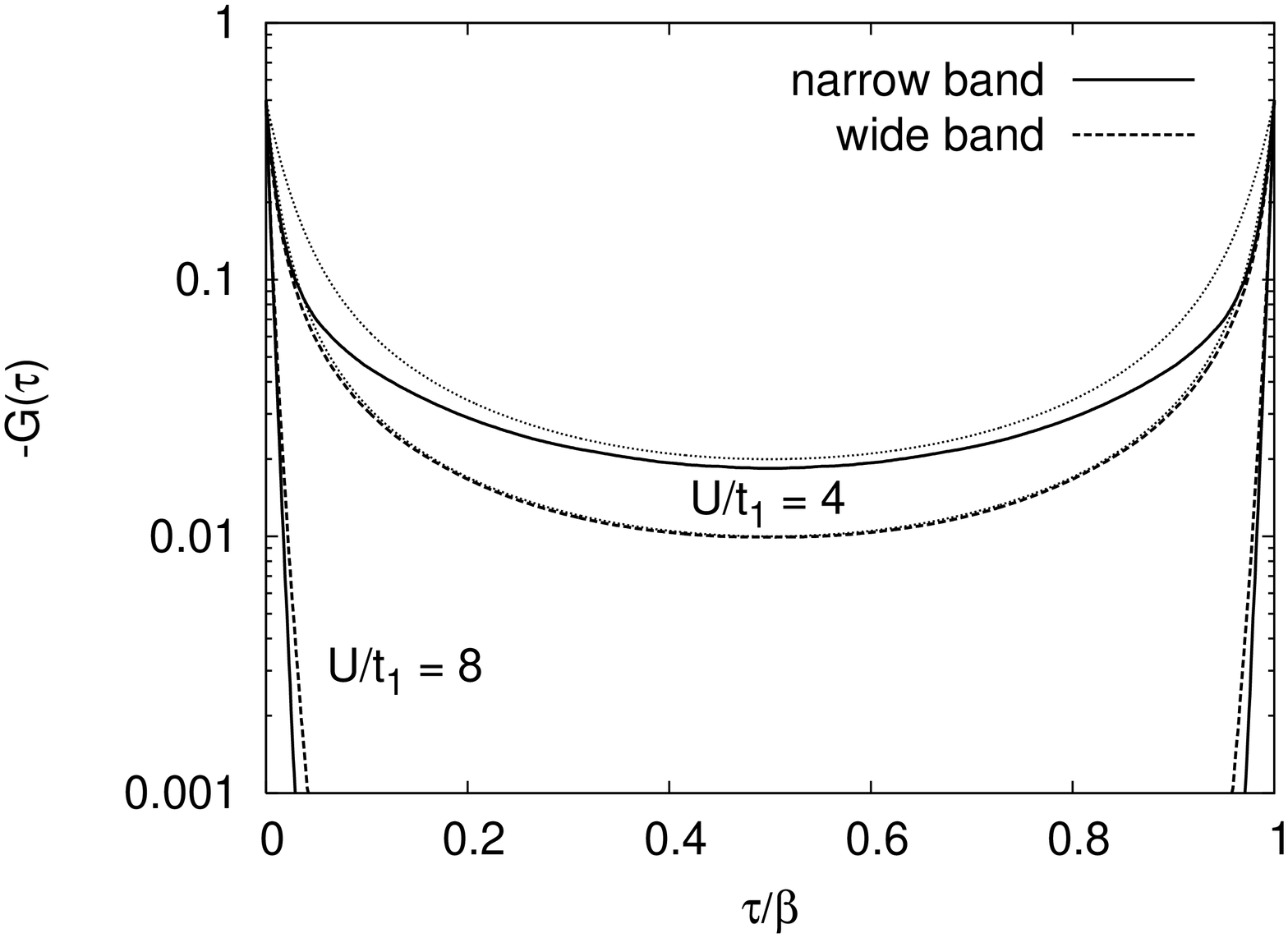}
\includegraphics [angle=-90, width=8.5cm] {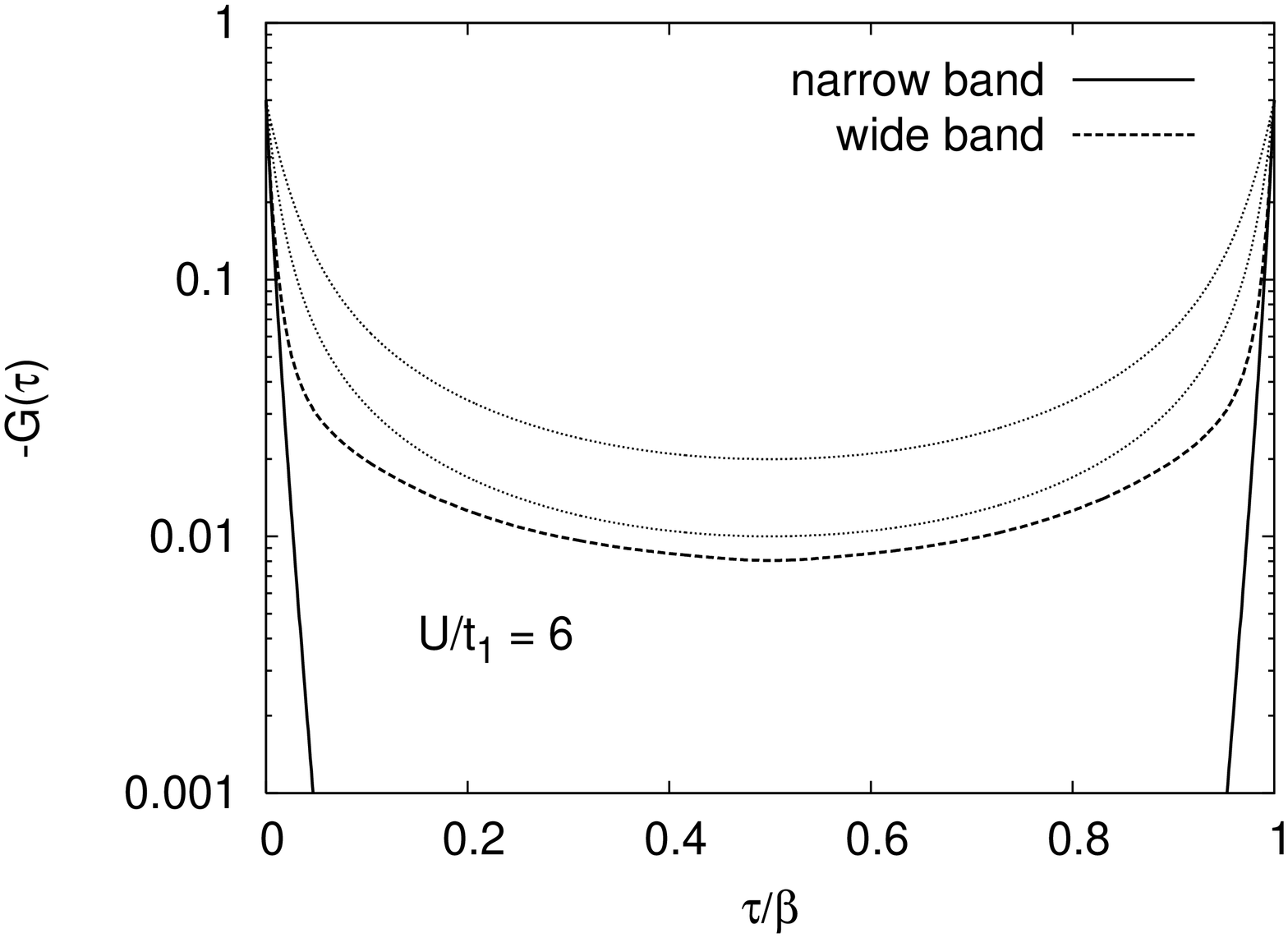}
\caption{Converged Green functions for the 2-orbital model with semicircular density of states at half-filling. The ratio of band widths is $t_2/t_1=2$ and the temperature $\beta t_1=50$. The exchange coupling is fixed as $J=U/4$. For $U/t_1=4$ (i.e. $U<U^c_1$), both bands are metallic, whereas for $U/t_1=8$ (i.e. $U>U^c_2$), both bands are insulating. For $U/t_1=6$, which lies in between $U^c_1$ and $U^c_2$, the narrow band is insulating, while the wide band is still metallic. Dotted lines show the non-interacting Green functions for $\beta t =50$ and $\beta t=100$.}
\label{orbital}
\end{figure}
Arita and Held \cite{Arita05} have recently used a new type of Hubbard-Stratonovich decomposition \cite{Sakai04}, which reduces the sign problem, and a projective QMC algorithm in their study of the two-orbital model. They found evidence for an orbital selective Mott transition in the presence of spin exchange, yet a single transition when merely the Ising component of the Hund's exchange was taken into account. Their estimate of $U^c_1$ was consistent with the value obtained in Ref.~\cite{Koga04}, while the projective QMC method did not allow to compute results at large enough couplings to estimate $U^c_2$. Other recent works \cite{Liebsch05, Inaba05, Sakai06} report the observation of two successive first order transitions and highlight the importance of taking the full Hund's coupling into account. It is therefore instructive to test our new algorithm on this example.

\begin{figure}[t]
\centering
\includegraphics [angle=-90, width=8.5cm] {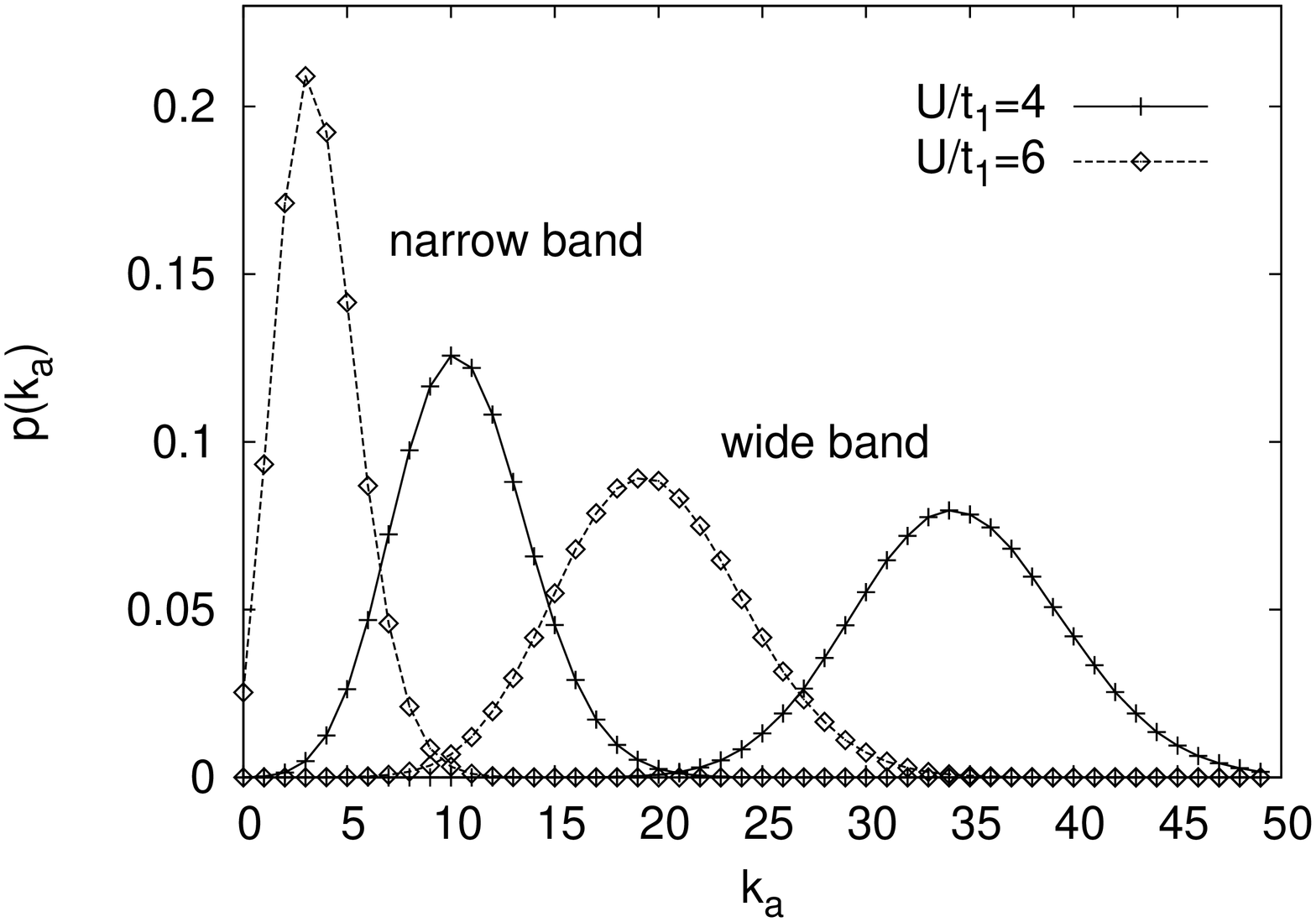}
\includegraphics [angle=-90, width=8.5cm] {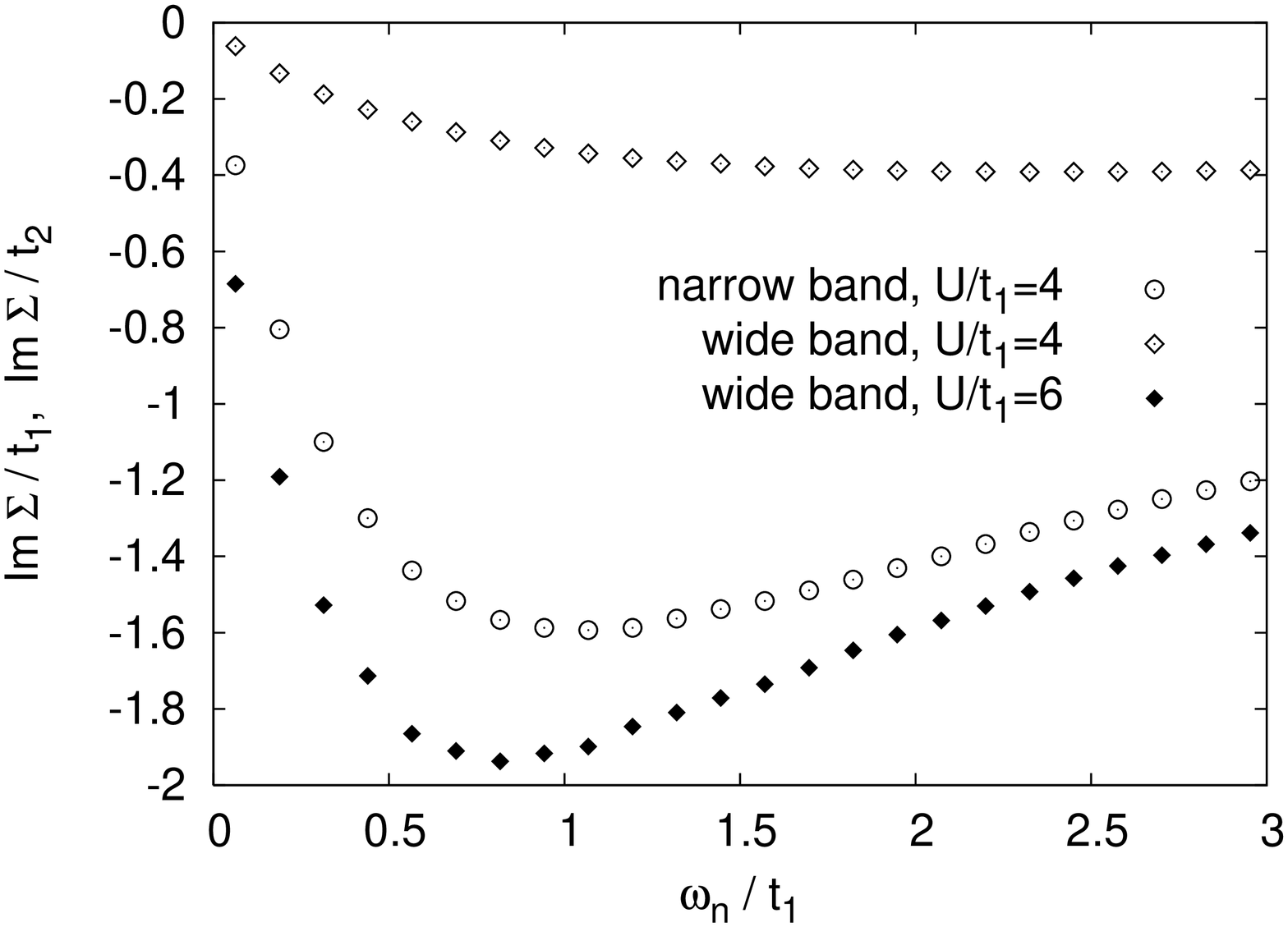}
\caption{Left panel: distribution of the orders $p(k_a)$, $a=1,2$ for $U/t_1=4$ and $U/t_1=6$. The average order is lower for the narrow band and decreases with increasing interaction strength. Right panel: imaginary part of the self-energies for the metallic states in the narrow and wide band, showing the strong correlations in the narrow band for $U/t_1=4$ and in the wide band for $U/t_1=6$. At $U/t_1=4$, the wide band is only weakly correlated.}
\label{2orbital_order}
\end{figure}

\begin{figure}[t]
\centering
\includegraphics [angle=-90, width=8.5cm] {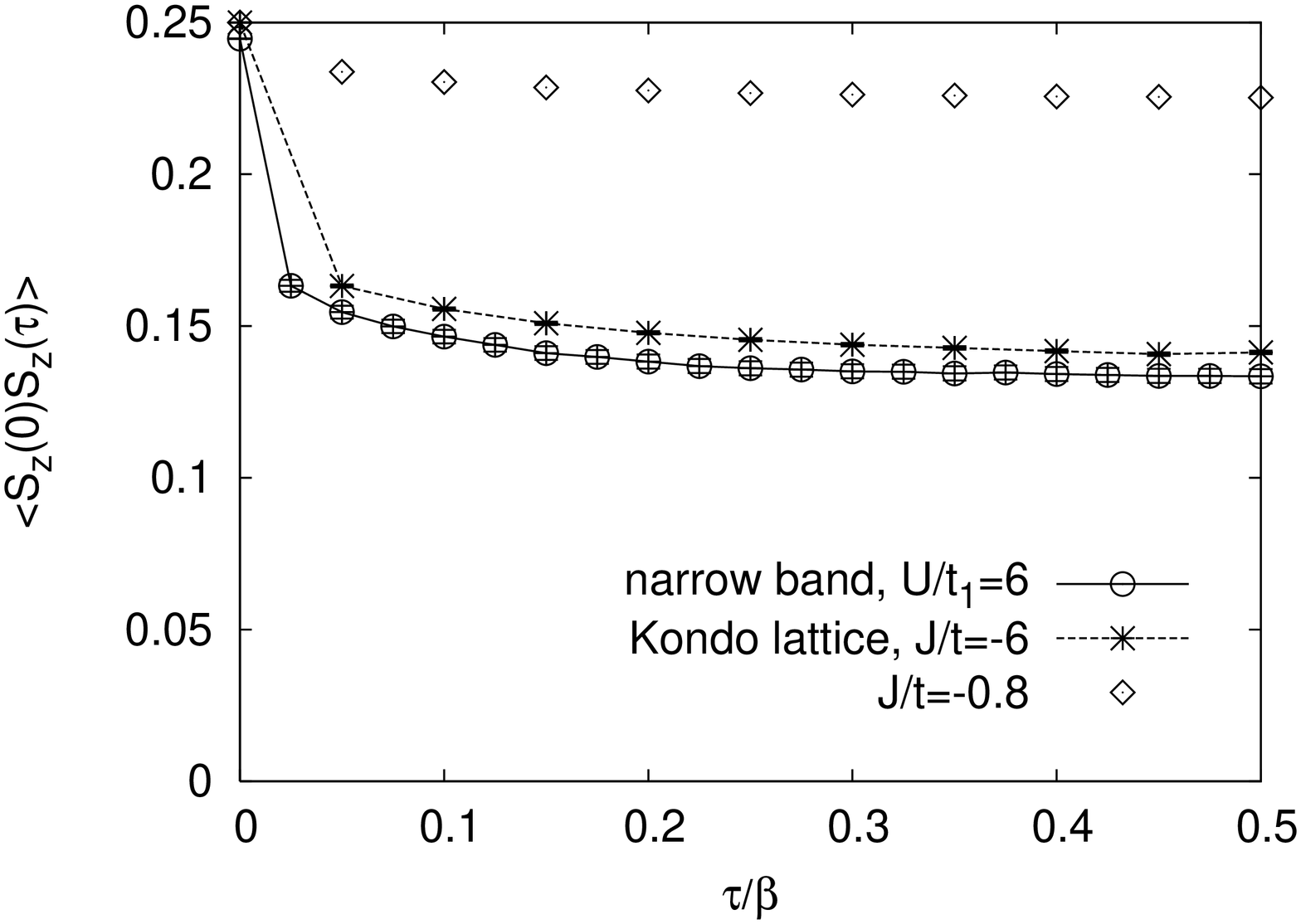}
\includegraphics [angle=-90, width=8.5cm] {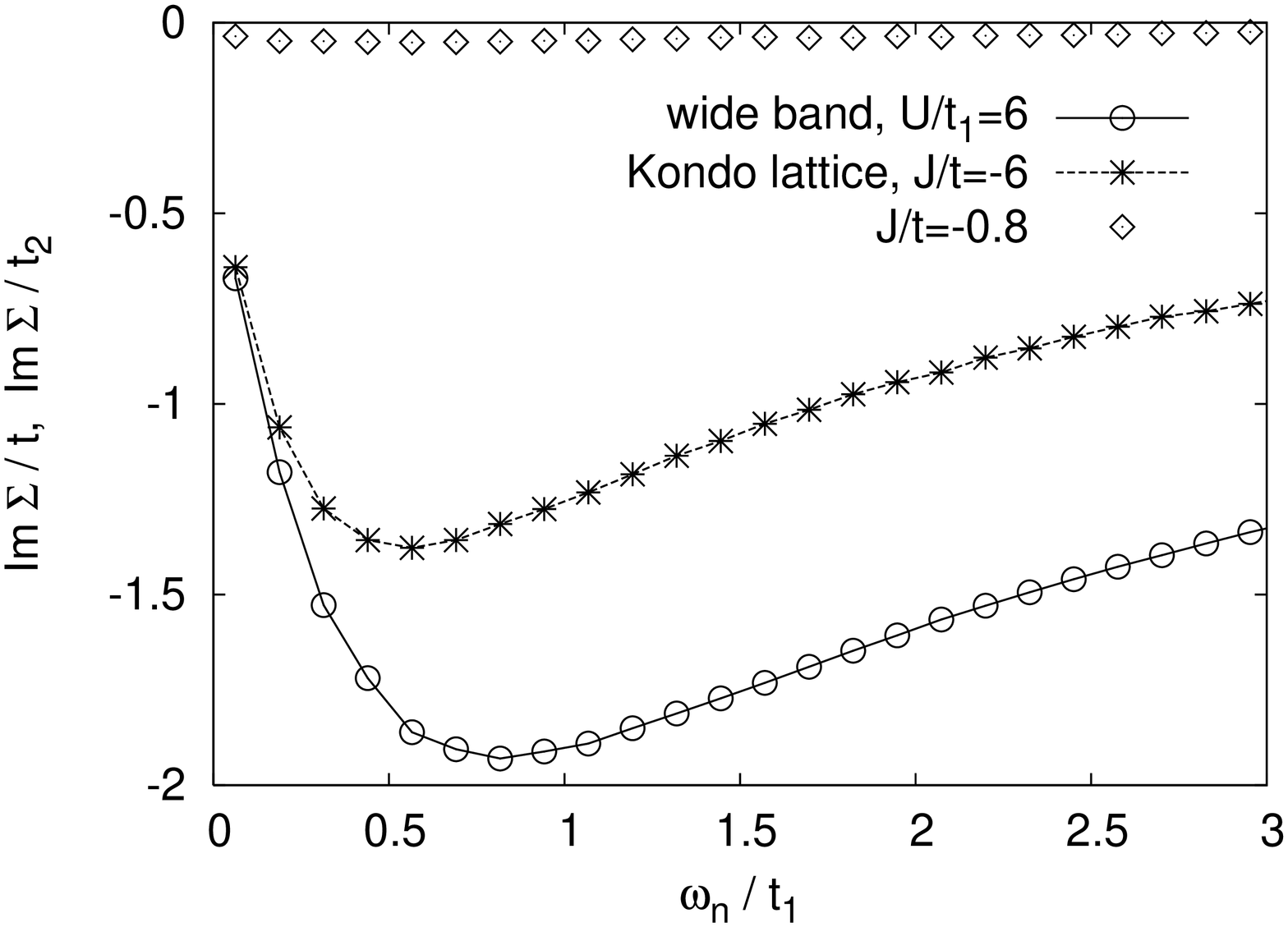}
\caption{Illustration of the relationship between an orbital selective state and the ferromagnetic kondo lattice model. The left figure shows the spin-spin correlation function in the narrow (insulating) band. The right figure shows the imaginary part of the self-energy for the wide (metallic) band. As in the ferromagnetic kondo lattice model, the correlation function in the insulating orbital saturates at large times, while the self-energy of the metallic band drops very rapidly as $\omega_n\rightarrow 0$.}
\label{2orbital_kondo}
\end{figure}

In Fig.~\ref{orbital} we show converged Green functions for $\beta t_1=50$, $J=U/4$ and $U/t_1=4$, 6, and 8. The chemical potential corresponds to half-filling and we average over spin up and down in each orbital. Our continuous-time algorithm does not suffer from any sign problem at these parameter values and the hybridization expansion approach allows us to access large interaction strengths. 
Fig.~\ref{2orbital_order} shows the distribution of orders in the two bands for the two coupling strengths $U/t_1=4$ and $U/t_1=6$. As expected, the average order is lower for the narrow band and the peaks shift to lower order as the interaction strength is increased.

At $U/t_1=4$, both bands are metallic, as can be seen in Fig.~\ref{orbital} from the long-time behavior 
of $G(\tau)$,
which is close to the non-interacting solutions, shown by the dotted lines. The shift in the narrow band indicates that the electrons, while still itinerant, 
are becoming more strongly correlated. At 
$U/t_1=8$, both bands are insulating as follows from the rapid (approximately exponential) drop of $G(\tau)$ to values much lower than the metallic solution. 
For $U/t_1=6$, Fig.~\ref{orbital} shows that $G(\tau)$ for the wide band is close to the non-interacting value, indicating metallic behavior, while $G(\tau)$ for the narrow band drops exponentially, indicating an insulating state.

The right hand panel of Fig.~\ref{2orbital_order} shows the self energies of the metallic bands for $U/t_1=4,6$. 
We see that for $U/t_1=4$ the
wide band is weakly correlated (self energy small compared to frequency and to the bandwidth) but the 
narrow band is strongly correlated. At $U/t_1=6$ the wide band is strongly correlated (self energy larger than frequency and indeed comparable to half the bandwidth), but similar to a fermi liquid
in the sense that $\Im m \Sigma(i\omega)\rightarrow 0$ as $\omega \rightarrow 0$. However, a more detailed analysis reveals interesting differences with conventional fermi liquid behavior. 

As noted by Biermann {\it et al.} \cite{Biermann05}, the insulating orbital is effectively a local moment, which is coupled to the metallic orbital by the exchange coupling $J$. The usual Hund's rules imply that the exchange is typically of ferromagnetic sign; thus in the orbital selective phase one might expect  
the model to map onto a ferromagnetic Kondo-Hubbard lattice,
with both an exchange coupling to a local moment and an on-site repulsion.  
Fig.~\ref{2orbital_kondo} shows that this is (at least qualitatively) indeed the case. The left hand panel plots the spin-spin correlation function of the insulating
orbital. The initial drop and saturation behavior characteristic of the ferromagnetic Kondo lattice model is evident.  (Note that because the orbital 1 can be empty or doubly occupied, the correlation function of the two orbital model  at $\tau=0$ is slightly less than 0.25.) The magnitude of the initial drop is surprisingly large. The plot was made for $U/t_1=6$ and $J/U=0.25$ implying $J/t_2=0.75$.
In the Kondo model, $J$'s of this magnitude lead to a much smaller decrease of $C_{SS}$ from its initial value. For comparison, we also plot results for $J/t=-0.8$ in Fig.~\ref{2orbital_kondo}, but as can be seen, an effective $J/t\approx -6$ is required to reproduce the 2-orbital results. Similarly, the right hand panel compares the calculated
self energy of the wide band to the Kondo lattice self energy corresponding to a $J/t$ chosen to approximately reproduce the drop in $C_{SS}$.
The qualitative behavior with a rapid decrease at low frequency is the same in both models, but the quantitative agreement is not good, suggesting that much of the self energy of the metallic band arises from the $U$, rather than from spin-dependent scattering due to the Kondo coupling to the localized orbital. We have not yet run simulations at low enough temperatures to test the occurrence of the power-law behavior in the self-energy, demonstrated in Fig.~\ref{qp_log} for the Kondo lattice model. 

Our results indicate that the ferromagnetic Kondo-Hubbard model exhibits an interesting interplay between the on-site repulsive interaction and the Kondo coupling, leading to a much larger effective exchange coupling than implied by the bare parameters. Further exploration of this physics is an important open issue.

\section{Conclusions}

We have presented a formalism which extends a previously proposed 
diagrammatic QMC method to wide classes of impurity models. The idea is to expand the partition function in the impurity-bath hybridization function, while treating the local part of the Hamiltonian exactly. The resulting matrix formalism 
allows an efficient simulation of models with reasonably small Hilbert spaces. We have demonstrated the usefulness of the new approach with simulation results for the Kondo lattice and two orbital models. In both cases, the simulations in physically interesting parameter regions do not suffer from a sign problem. 

The new formalism opens up wide classes of questions for  investigation. Systemtic investigations of quasiparticle and
magnetic properties
of orbital selective Mott phases are now possible. We have provided direct calculations which
support the conjecture  of Biermann {\it et al.} \cite{Biermann05} that the orbital selective Mott phase is in some qualitative
sense described by an effective ferromagnetic 
Kondo lattice model, and we have further demonstrated that the Coulomb correlations in this phase
play a very important role, leading 
to an effective coupling much larger than expected from the basic scales of the model. Concerning the Kondo lattice
model, we have shown by comparing the ferromagentically and antiferromagnetically coupled
cases that the renormalizations familiar from the one-impurity problem survive and 
have pronounced effects on the lattice problem, even at interaction scales of the order of unity. For example, the Neel temperature of the $|J|=1$ models differ considerably in the ferromagnetic and antiferromagnetic cases, which we believe is a result of the opposite renormalization of $J$ in the two cases. For the  ferromagnetic  Kondo lattice 
model, we have discovered an unusual power law renormalization of the electron self energy which we propose
is related to the density of states renormalization associated with the $J$-driven metal-insulator transition. Further
investigation of this transition will be a fruitful subject for future research.  For the antiferromagnetically
coupled model we have located the Kondo-insulator to antiferromagnet transition and shown that the variation of the magnetization near the transition point is extremely rapid.  
    
Our method is from a conceptual and technical point of view appealing, 
because it does not require a double expansion in both the hybridization and the exchange couplings. 
The algorithm leads to manageable perturbation orders and, in the models studied so far, to 
an undetectably small sign problem in relevant regions of parameter space.   
In the presence of exchange processes, however, one has to compute the trace over all 
basis states in Eq.~(\ref{detailed_balance}) explicitly.  Because the Hilbert space
grows exponentially with the number of orbitals, a straightforward application of the procedure
introduced here becomes impractical for large impurity problems (a four site 
Hubbard cluster with $256$ basis states seems about the largest system one might want to consider).  

We see two possible ways to approach this problem. A straight-forward alternative is the above mentioned double expansion, which allows one to return to the economical segment picture \cite{Werner06} to represent the configurations and to devise efficient Monte Carlo moves which are compatible with the constraints of the model. Since the exchange couplings in many relevant models tend to be weak, the increase in the perturbation orders should still be manageable. What will happen to the sign problem remains to be seen. 

Another approach is based on the observation that most of the states in the exponentially large Hilbert space are of very high energy and are therefore not directly relevant to the physics. An important issue for future research is the development of  ``effective action" methods which will allow the elimination of high energy states, reducing the problem to one with a much smaller Hilbert space, to which the matrix formalism can be directly applied. 

\acknowledgments

Support from NSF DMR 0431350 is greatfully acknowledged. The calculations have been performed on the Hreidar Beowulf cluster at ETH Z\"urich, using the ALPS library \cite{ALPS}. We thank M.~Troyer for the generous allocation of computer time, C. Lin for
the classical spin data in Figs.~\ref{magnetic_ferro} and \ref{magnetic_antiferro} and A.~Georges, S.~Biermann, L.~de'~Medici, I.~Milat, M.~Sigrist and S.~Okamoto for helpful discussions.


\begin{thebibliography}{99}
\bibitem{Georges96} A. Georges, G. Kotliar, W. Krauth and M. J. Rozenberg, Rev. Mod. Phys. {\bf 68}, 13 (1996).
\bibitem{Jarrell} M. H. Hettler, A. N. Tahvildar-Zadeh, M. Jarrell, T. Pruschke, and H. R. Krishnamurthy, Phys. Rev. {\bf B58}, 7475 (1998); T. Maier, M. Jarrell, T. Pruschke, and M. H. Hettler 
Rev. Mod. Phys. {\bf 77}, 1027-1080 (2005).
\bibitem{CDMFT} G. Kotliar, S. Y. Savrasov, G. Palsson, and G. Biroli Phys. Rev. Lett. {\bf 87}, 186401 (2001).
\bibitem{Fuhrmann04} S. Okamoto, A. J. Millis, H. Monien, and A. Fuhrmann Phys. Rev. {\bf B68}, 195121 (2003).
\bibitem{Hirsch86} J. E. Hirsch and R. M. Fye, Phys. Rev. Lett. {\bf 56}, 2521 (1986). 
\bibitem{Caffarel94} M. Caffarel and W. Krauth, Phys. Rev. Lett. {\bf 72},1545 (1994).
\bibitem{Capone02} M. Capone, M. Fabrizio, C. Castellani, and E. Tosatti, Science 296, 2364 (2002).
\bibitem{Kyung06} B. Kyung, A.-M. S. Tremblay, cond-mar/0604377.
\bibitem{Sakai04} S. Sakai, R. Arita and H. Aoki, Phys. Rev. \textbf{B70}, 172504 (2004).
\bibitem{Arita05} R. Arita and K. Held, Phys. Rev. \textbf{B72}, 201102(R) (2005). 
\bibitem{Koga05} A. Koga, N. Kawakami, T. M. Rice and M. Sigrist, Phys. Rev. \textbf{B72}, 045128 (2005).
\bibitem{Rubtsov05} A. N. Rubtsov, V. V. Savkin and A. I. Lichtenstein, Phys. Rev. {\bf B72}, 035122 (2005).
\bibitem{Werner06} P. Werner, A. Comanac, L. de' Medici, M. Troyer and A. J. Millis, Phys. Rev. Lett. \textbf{97}, 076405 (2006). 
\bibitem{Savkin05} V. V. Savkin, A. N. Rubtsov, M. I. Katsnelson and A. I. Lichtenstein, Phys. Rev. Lett. \textbf{94}, 026402 (2005).
\bibitem{Rombouts99} S. M. A. Rombouts, K. Heyde and N. Jachowicz, Phys. Rev. Lett. \textbf{82}, 4155 (1999).
\bibitem{Sakai06} S. Sakai, R. Arita, K. Held and H. Aoki, cond-mat/0605526.
\bibitem{Chattopadhyay01} A. Chattopadhyay, A. J. Millis and S. Das Sarma, Physical Review \textbf{B64}, 012416/1-4, (2001).
\bibitem{kondo_gap} V. Dorin and P. Schlottmann, Phys. Rev. {\bf B46}, 10800Ð10807 (1992).
\bibitem{Doniach77} S. Doniach, Physica {\bf B91}, 231 (1977).
\bibitem{Kienert06} J. Kienert and W. Nolting, cond-mat/0606485.
\bibitem{Biermann05} S. Biermann, L. de' Medici and A. Georges, Phys. Rev. Lett. \textbf{95}, 206401 (2005). 
\bibitem{Lin06} C. Lin, private communication.
\bibitem{Koga04} A. Koga, N. Kawakami, T. M. Rice and M. Sigrist, Phys. Rev. Lett. \textbf{92}, 216402 (2004).
\bibitem{Liebsch03} A. Liebsch, Phys. Rev. Lett. \textbf{91}, 226401 (2003).
\bibitem{Liebsch05} A. Liebsch, Phys. Rev. Lett. \textbf{95}, 116402 (2005).
\bibitem{Inaba05} K. Inaba, A. Koga, S. Suga and N. Kawakami, J. Phys. Soc. Jpn. \textbf{74}, 2393 (2005).
\bibitem{ALPS} M. Troyer {\it et al.}, Lecture Notes in Computer Science {\bf 1505}, 191 (1998); F. Alet \textit{et al.}, J. Phys. Soc. Jpn. Suppl. {\bf 74}, 30 (2005); \url{http://alps.comp-phys.org/} .
\end{thebibliography}
\end{document}